\documentclass[twocolumn]{aastex61}

\usepackage{amsmath}
\usepackage{graphicx}

\usepackage{epsfig}
\usepackage{wasysym}
\usepackage{lipsum}
\usepackage{mathrsfs}
\usepackage{float}
\usepackage{rotating}
\usepackage{graphicx}
\usepackage{epstopdf}
\usepackage{array}
\newcolumntype{P}[1]{>{\centering\arraybackslash}p{#1}}
\newcolumntype{M}[1]{>{\centering\arraybackslash}m{#1}}
\newcounter{chem}
\newcounter{temp}

\usepackage{amsmath}
\usepackage{mhchem}

\providecommand{\e}[1]{\ensuremath{\times 10^{#1}}}

\usepackage{color}

\usepackage{gensymb}

\usepackage{filecontents}
\usepackage{filecontents}

\begin{document}

\shorttitle{3D Chemistry-Climate near the IHZ around M-stars}
\shortauthors{Chen, H., Wolf, E. T., et al.}

\title{Habitability and Spectroscopic Observability of Warm M-dwarf Exoplanets \\ Evaluated with a 3D Chemistry-Climate Model}

\author[0000-0003-1995-1351]{Howard Chen}

\affil{Department of Earth and Planetary Sciences, Northwestern University, Evanston, IL 60202, USA}
\affil{Center for Interdisciplinary Exploration \& Research in Astrophysics (CIERA), Evanston, IL 60202, USA}

\author[0000-0002-7188-1648]{Eric T. Wolf}

\affil{Laboratory for Atmospheric and Space Physics, Department of Atmospheric and Oceanic Sciences, University of Colorado Boulder, Boulder, CO 80309, USA}
\affil{NASA Astrobiology Institute Virtual Planetary Laboratory, Seattle, WA 98194, USA}

\author[0000-0002-4142-1800]{Zhuchang Zhan}
\affil{Massachusetts Institute of Technology, Department of Earth, Atmospheric, and Planetary Sciences, Cambridge, MA 02138, USA}

\author[0000-0002-2065-4517]{Daniel E. Horton}

\affil{Department of Earth and Planetary Sciences, Northwestern University, Evanston, IL 60202, USA}
\affil{Center for Interdisciplinary Exploration \& Research in Astrophysics (CIERA), Evanston, IL 60202, USA}

\begin{abstract}
Planets residing in circumstellar habitable zones (CHZs) offer our best opportunities to test hypotheses of life's potential pervasiveness and complexity. Constraining the precise boundaries of habitability and its observational discriminants is critical to maximizing our chances at remote life detection with future instruments. Conventionally, calculations of the inner edge of the habitable zone (IHZ) have been performed using both 1D radiative-convective and 3D general circulation models. However, these models lack interactive 3D chemistry and do not resolve the mesosphere and lower thermosphere (MLT) region of the upper atmosphere. Here we employ a 3D high-top chemistry-climate model (CCM) to simulate the atmospheres of synchronously-rotating planets orbiting at the inner edge of habitable zones of K- and M-dwarf stars (between $T_{\rm eff} =$ 2600 K and 4000 K). While our IHZ climate predictions are in good agreement with GCM studies, we find noteworthy departures in simulated ozone and HO$_{\rm x}$ photochemistry. For instance, climates around inactive stars do not typically enter the classical moist greenhouse regime even with high ($\ga 10^{-3}$ mol mol$^{-1}$) stratospheric water vapor mixing ratios, which suggests that planets around inactive M-stars may only experience minor water-loss over geologically significant timescales.  In addition, we find much thinner ozone layers on potentially habitable moist greenhouse atmospheres, as ozone experiences rapid destruction via reaction with hydrogen oxide radicals. Using our CCM results as inputs, our simulated transmission spectra show that both water vapor and ozone features could be detectable by instruments NIRSpec and MIRI LRS onboard the James Webb Space Telescope.
\end{abstract}


\correspondingauthor{Howard Chen, NASA Future Investigator}
\email{howard@earth.northwestern.edu}

\section{Introduction} 
\label{sec:intro} 

For the first time in human history, it is possible to find and characterize nearby rocky and potentially habitable worlds. Recent discoveries of Proxima Centauri b, the TRAPPIST-1 system, and LHS 1140b \citep{AngladaEt2016NATURE,GillonEt2017NATURE,DittMannEt2017NATURE} show that remote examination of small rocky planets is within reach. Terrestrial planets, such as these, are expected to be common (${\sim}15\%$) in circumstellar habitable zones (CHZs) of low-mass stars \citep{TarterEt2007,Dressing+Charbonneau2015ApJ} $-$ systems that are especially amenable to spectroscopic observation due to their high transit frequencies, low star-to-planet brightness contrasts, prolonged main sequence lifetimes, and abundance in both the solar neighborhood and the projected Transiting Exoplanet Survey Satellite (TESS) sample \citep{HenryEt2006AJ,BarclayEt2018ApJS}. In fact, TESS has already found small and Earth-sized planets transiting cool stars (e.g., \citealt{VanderspekEt2019ApJL,DragomirEt2019ApJL}). Upcoming characterization efforts by the James Webb Space Telescope, ground-based 30-meter extremely large telescopes, and direct imaging missions will likely attempt to detect habitability indicators (e.g., N$_2$, H$_2$O$_v$) and/or biosignatures (O$_2$, O$_3$, CH$_4$, N$_2$O, CO$_2$; \citealt{SaganEt1993NATURE}) on these K- and M-dwarf systems. Indeed, atmospheric characterization of increasingly smaller planets ($R_p \la 4~R_\oplus$) is already underway (e.g., \citealt{WakefordEt2017SCI,BennekeEt2019NAtAstro,BennekeEt2019arXiv}).

Future target selection and characterization efforts will benefit from improved understanding of and constraints on CHZ boundaries. Earliest estimates of the CHZ made use of energy balance models (EBMs) \citep{Hart1979Icarus}, which established the dependence of HZ widths on stellar spectral type. Follow on studies, using 1D radiative-convective models, identified two boundaries of the inner habitable zone: one defined by the onset of a water-enriched stratosphere and another defined by a radiative equilibrium threshold \citep{KastingEt1984Icarus,Kasting1988Icarus}. These early simulations assumed a fully saturated troposphere with a fixed moist adiabatic lapse rate and static clouds. Subsequently, \citet{KastingEt2015ApJL} found that as the absorbed stellar flux increases, the stratosphere moistens and warms significantly which could allow water vapor to efficiently escape to space. Other studies, also using 1D models, provided additional insights, finding for example, that CHZ widths could change according to atmospheric composition and/or atmospheric pressure (e.g., \citealt{VladiloEt2013ApJ,ZsomEt2013ApJ,Ramirez+Kaltenegger2017ApJL}).

In more recent years, idealized and state-of-the-art estimates of the CHZ have utilized 3D general circulation models (GCMs) to place physics-based constraints on CHZ boundaries (e.g., \citealt{AbeEt2011AsBio,YangEt2014ApJL,WayEt2016GRL}). GCM predictions improved upon 1D model projections by way of explicit simulation of large-scale circulation and key climate system feedbacks. For instance, incorporation of atmospheric dynamics into models of slowly-rotating planets resulted in climatic behaviors that can only be resolved in 3D, e.g., substellar cloud formation and convergence caused by changes in the Coriolis force \citep{YangEt2013ApJL,KopparapuEt2016ApJ,WayEt2018ApJS}. Follow-up studies, using similar GCMs, found that habitable planets around M-dwarf stars have moist stratospheres despite mild global mean surface temperatures (e.g., \citealt{FujiiEt2017ApJ,KopparapuEt2017ApJ}). These results stood in contrast to previous inverse modeling approaches with 1D radiative-convective models (e.g., \citealt{Kasting1988Icarus}), where a surface temperature of 340 K was deemed the threshold for the classical ``moist greenhouse" regime.

Despite these advances, exoplanet GCM studies have traditionally not accounted for photochemical and atmospheric chemistry-climate interactions $-$ components recently found by both 1D \citep{LincowskiEt2018ApJ,KozakisEt2018ApJ} and 3D models \citep{ChenEt2018ApJL} to be critical for habitability and biosignature prediction. The addition of photochemistry to prognostic atmospheric models allows for interactions between high energy photons and gaseous molecules. This often leads to the breaking of molecular bonds and creation of free radicals and ions, which have significant impacts on atmospheric composition and associated habitability. Determination of water loss, in particular, requires knowledge of where water vapor photodissociation occurs in the mesosphere and lower thermosphere (MLT), and is dependent on dynamical, photochemical, and radiative processes. To simulate the speciation, reaction, and transport of various gaseous constituents (e.g., H$_2$O$_v$) and their photochemical byproducts (e.g., H, H$_2$), coupled 3D chemistry-climate models (CCMs) are needed. CCMs are also able to simulate photochemically important species such as ozone, allowing for prognostic assessments of chemistry-climate system feedbacks. As ozone is primarily derived from molecular oxygen, prognostic ozone calculations enable consideration of O$_2$-rich atmospheres with active oxygenating photosynthesis on the surface. Lastly, the large number of chemical species calculated by CCMs provide a rich tapestry for calculating transmission spectra, compared to the simplified atmospheric compositions generally considered in GCMs.

Earlier climate models of tidally-locked Earth-like planets adopted vertical resolutions of 15-25 layers with equally spaced 30-50 mbar levels \citep{JoshiEt1997Icarus,Merlis+Schneider2010,EdsonEt2011Icar}. While this setup fully resolved the general structure of the troposphere where the majority of weather takes place \citep{Held+Suarez1994}, it neglected the upper stratosphere and thermosphere. Critically, interactive simulation of photochemistry and atmospheric chemistry requires a high model-top (i.e., a model whose atmosphere reaches into the mid-thermosphere (${\sim} 150$ km)), as highly energetic photons initially and primarily interact with a planet’s upper atmosphere. While most radiatively active species are stable against dissociation in the troposphere, species become vulnerable to photolysis above the tropopause and to photoionization above the stratopause (${\sim}1$ mbar) and mesopause (${\sim}10^{-2}$ mbar). Apart from key dissociative processes in the MLT region, high-top atmospheric dynamics are also important as they influence the transport of gaseous molecules. Vertical velocity in the vicinity of the tropopause is not an isolated process, but influenced by momentum sources in the stratosphere and lower thermosphere \citep{HoltonEt1995}. Further, mean meridional circulation of the lower stratosphere, which can affect the distribution of chemical tracers, is driven primarily by the drag provided by planetary and gravity wave momentum deposition in the stratosphere and mesosphere \citep{HaynesEt1991,HoltonEt1995,Romps+Kuang2009GRL}. A high model-top is therefore essential to simulate chemical interactions and their associated dynamical processes in the MLT region.

Based on conventional theory, endmembers of habitability are represented by (i) CO$_2$-rich icehouse climates at the outer edge of the habitable zone (OHZ; e.g., \citealt{Paradise+Menou2017ApJ}) and (ii) moist greenhouse climates at the inner edge of the habitable zone (IHZ; e.g., \citealt{KopparapuEt2016ApJ}).  In this study, we use a 3D high-top CCM to investigate the latter regime, namely the moist greenhouse limits of IHZ planets orbiting M-dwarf stars. The paper is organized as follows: In Section 2, we describe our model and experimental setup. In Section 3, we present and analyze our results. Section 4 discusses implications of our results, caveats, and relevance to observations. Finally, Section 5 summarizes key findings, provides concluding remarks, and suggests next step.

\begin{deluxetable*}{ccccccccccc}




\tablecaption{Summary of Experiments Performed in this Study}

\tablenum{1}

\tablehead{\colhead{Name} & \colhead{$T_{\rm eff}$} & \colhead{$L_*$} & \colhead{$M_*$} & \colhead{$F_p$} & \colhead{$P_p$} & \colhead{T$_s$} & \colhead{H$_2$O$_v$} & \colhead{H} & \colhead{$T_{\rm th}$} & \colhead{Bond Albedo}\\ 
\colhead{} & \colhead{(K)} & \colhead{($L_\odot$)} & \colhead{($M_\odot$)} & \colhead{($F_\oplus$)} & \colhead{Earth days} & \colhead{(K)} & \colhead{(mol mol$^{-1}$)} & \colhead{(mol mol$^{-1}$)} & \colhead{(K)} & \colhead{} }

\startdata
09F26T & 2600 & 0.000501 & 0.0886 & 0.9 & 4.43 & 268.36 & 2.95e-06 & 4.01e-06 & 341.90 &   0.32  \\
{\bf 10F26T} & 2600 & 0.000501 & 0.0886 & 1.0 & 4.11 & 282.39 & 4.22e-06 & 3.68e-06 & 314.14  &  0.23 \\
11F26T & 2600 & 0.000501 & 0.0886 & 1.1 & 3.82 & runaway & - & - & - &-  \\
10F30T & 3000 & 0.00183 & 0.143 & 1.0 & 8.53 & 270.39 & 4.99e-06 & 3.70e-06 & 337.27 & 0.38 \\
{\bf 11F30T} & 3000 & 0.00183 & 0.143 & 1.1 & 7.91 & 278.76 & 7.52e-06 & 3.87e-06 & 329.97 & 0.33  \\
12F30T & 3000 & 0.00183 & 0.143 & 1.2 & 7.41 & runaway & - & - & -&-  \\
10F33T & 3300 & 0.00972 & 0.249 & 1.0 & 23.13 & 266.73 & 4.63e-06 & 3.70e-06 & 333.57 &  0.45  \\
15F33T & 3300 & 0.00972 & 0.249 & 1.5 & 16.23 & 284.33 & 2.61e-04 & 5.47e-06 & 342.75 &  0.46 \\
{\bf 16F33T} & 3300 & 0.00972 & 0.249 & 1.6 & 15.46 & 297.75 & 2.05e-03 & 1.22e-04 & 343.39  & 0.47    \\
17F33T & 3300 & 0.00972 & 0.249 & 1.7 & 14.77 & runaway & - & - & - & -\\                                      
10F40T & 4000 & 0.0878 & 0.628 & 1.0 & 69.39 & 265.76 & 2.66e-06 & 3.69e-06 & 319.54 & 0.48  \\
17F40T & 4000 & 0.0878 & 0.628 & 1.7 & 47.64 & 282.88 & 3.40e-04 & 6.23e-06 & 330.04 &  0.52   \\
{\bf 19F40T} & 4000 & 0.0878 & 0.628 & 1.9 & 43.87 & 301.41 & 1.18e-02 & 7.27e-05 & 342.75 & 0.55   \\
20F40T & 4000 & 0.0878 & 0.628 & 2.0 & 42.22 & runaway & - & - & - &- \\
19FSolarUV$^a$ & 4000 & 0.0878 & 0.628 & 1.9 & 43.87 &  303.38 & 1.30e-02 & 1.05e-03 & 409.62 & 0.51 \\
19FADLeoUV$^b$ & 4000 & 0.0878 & 0.628 & 1.9 & 43.87 &  315.29 & 9.65e-02 & 7.45e-02 & 382.15  & 0.45 \\
\enddata


\tablecomments{Summary of model initial conditions and simulated results pertinent to habitability. Values presented are: effective temperature of host star ($T_{\rm eff}$), luminosity of host star in solar units ($L_*$), mass of host star in solar units ($M_*$), incident flux on the planet's dayside relative to Earth's annual average incident flux (i.e., 1362 W m$^{-2}$; $F_p$), rotation period of the planet in Earth days ($P_p$), simulated global-mean surface temperature (T$_s$), simulated water vapor mixing ratio at 0.1 mbar (H$_2$O$_v$), simulated model-top thermospheric hydrogen mixing ratio (H), and simulated thermospheric temperature at 100 km altitude ($T_{\rm th}$). Bolded experiments indicate ``IHZ limit" simulations, which are the focus of discussion in Section 3.2. $^a$Simulation with input stellar SED modified by swapping out the UV bands ($\lambda < 300$ nm) of the fiducial star and replacing it with $2 \times$ Solar UV from \citet{lean1995reconstruction}. $^b$AD Leonis UV data from \citet{SeguraEt2005AsBio}.}


\end{deluxetable*}

\section{Model Description \& Numerical Setup}
\label{sec:method}

We employ the National Center for Atmospheric Research (NCAR) Whole Atmosphere Community Climate Model (WACCM) to investigate the putative atmospheres of rocky exoplanets. WACCM is a 3D global CCM that simulates interactions of atmospheric chemistry, radiation, thermodynamics, and dynamics. We set the Community Atmosphere Model v4 (CAM4) as the atmosphere component of WACCM. CAM4 uses native Community Atmospheric Model Radiative Transfer (CAMRT) radiation scheme \citep{kiehl1983co2}, the Hack scheme for shallow convection \citep{Hack1994JGR}, the Zhang-McFarlane scheme for deep convection \citep{zhang1995sensitivity}, and the Rasch-Kristjansson (RK) scheme for condensation, evaporation and precipitation \citep{zhang2003modified}. For a complete model description see \citet{neale2010description} and \citep{MarshEt2013JGR}.

WACCM includes an active hydrological cycle and prognostic photochemistry and atmospheric chemistry reaction networks. The chemical model is version 3 of the Modules for Ozone and Related Chemical Tracers (MOZART) chemical transport model \citep{KinnisonEt2007JGR}. The module resolves 58 gas phase species including neutral and ion constituents linked by 217 chemical and photolytic reactions. The land model is a diagnostic version of the Community Land Model v4 with the 1850 control setup including prescribed surface albedo, surface CO$_2$, vegetation, and forced cold start. The oceanic component is a 30-meter deep thermodynamic slab model with zero dynamical heat transport and no sea-ice. Even though a fully dynamic ocean is ideal, \citet{YangEt2019ApJa} recently demonstrated that the inclusion of such does not significantly impact the climate of moist greenhouse IHZ planets. Furthermore, the presence of significant North-South oriented continents minimizes the effects of ocean heat transport on tidally locked worlds \citep{YangEt2019ApJa,DelGenioEt2019AsBio}.

WACCM includes a number of improved and expanded high-top atmospheric physics and chemical components. Processes in the MLT region are based on the thermosphere-ionosphere-mesosphere electrodynamics (TIME) GCM \citep{RobleEt1994GRL}. Key processes included are: neutral and high-top ion chemistry (ion drag, auroral processes, and solar proton events) and their associated heating reactions. In this study, we do not use prognostic ion chemistry (e.g., WACCM-D; Verronen et al. 2016) for the sake of computational efficiency. In terms of atmospheric dynamics, WACCM allows the emergence of gravity waves (important for governing large-scale flow patterns and chemical transport) by orographic sources, convective overturning, or strong velocity shears \citep{neale2010description}. As we assume Earth-like topography in all our simulations, orography may also provide a means of direct forcing on planetary scale Rossby waves, which act to increase the asymmetry of atmospheric circulation and turbulent flow (the absence of topography such as on an idealized aquaplanet would minimize this effect and thus induce greater circulation symmetry). Topography also drives gravity waves which deposit energy into the mesosphere, affecting its temperature structure and circulation \citep{BardeenEt2010JGR}. Molecular diffusion via gravitational separation of different molecular constituents \citep{BanksET1973} is an extension to the nominal diffusion parameterization in CAM4. Below 65 km (local minimum in shortwave heating and longwave cooling), WACCM retains CAM4’s radiation scheme. Above 65 km, WACCM expands upon both longwave (LW) and shortwave (SW) radiative parameterizations from those of CAM3 and CAM4 \citep{CollinsEt2006}. WACCM uses thermodynamic equilibrium (LTE) and non-LTE heating and cooling rates in the extreme ultraviolet (EUV) and infrared (IR)\citep{FomichevE1998JGR}. In the SW (0.05 nm to 100 $\mu$m; \citealt{Lean2000GRL,Solomon+Qian2005JGR}), radiative heating and cooling are sourced from photon absorption, as well as photolytic and photochemical reactions. To our knowledge, we are the first to apply these models and modules in the context of exoplanets.

Earth's atmospheric structure is typically defined using the vertical temperature gradient. In WACCM simulations the atmospheric structure is dependent on the atmospheric gases, planetary rotation period, and bolometric stellar flux $-$ thus the simulated vertical temperature gradient is different from that of Earth. However, to facilitate comparison, we refer to simulated atmospheric layers using the typical pressure levels of Earth’s atmosphere: That is, troposphere refers to regions extending from the surface to 200 mbar, the stratosphere 200 mbar to 1 mbar, the mesosphere 1 mbar to 0.001 mbar, and the thermosphere 0.001 mbar to $5\e{-6}$ mbar. The latter two layers comprise the so-called MLT region (see Section 1). Specifically, the mesosphere extends from ${\sim}1$ mbar to the mesopause (roughly the homopause, at ${\sim}0.001$ mbar), where temperature minima caused by CO$_2$ radiative cooling are typically found. Above the mesosphere, the thermosphere encompasses the heterosphere zone, in which diffusion plays an increasingly greater role and the chemical composition of the atmosphere varies in accordance with the atomic and molecular mass of each species. This region (i.e., the thermosphere) extends from the mesopause to model-top at pressures of $5.1 \e{-6}$ hPa (${\sim}145$ km). Shifts in atmospheric structure occur and depend on the planetary rotation period and the bolometric stellar flux.

We configure the described model components to simulate the atmospheres of tidally-locked (trapped in 1:1 spin-orbit resonance) planets across a range of stellar spectral energy distributions, bolometric stellar fluxes, and planetary rotation periods (Table 1). We construct stellar spectral energy distributions (SEDs) using the PHOENIX synthetic spectra code \citep{HusserEt2013A&A} assuming stellar metallicities of [Fe/H] = 0.0, alpha- enhancements of [$\alpha$/M] = 0.0, surface gravities log g = 4.5, and stellar effective temperatures ($T_{\rm eff}$) of 2600 K (TRAPPIST- 1-like), 3000 K (Proxima Centauri-like), 3300 K (AD Leo-like), and 4000 K (late K-dwarf). All synthetic stellar SEDs are assumed to be in states of quiescence. The corresponding rotation periods obeys Kepler's 3rd law, as in \citet{KopparapuEt2016ApJ}, which is given by:

\begin{equation}
    P_{\rm years} = \left[ \left( \frac{L_*/L_\odot}{F_p/F_\oplus}\right)^{3/4}\right] \left(M_*/M_\odot \right)^{1/2}
\end{equation}

\noindent where $L/L_\odot$ is the stellar luminosity in solar units, $F_p/F_\oplus$ is the incident stellar flux in units of present-Earth flux (1360 W m$^{-2}$), and $M/M_\odot$ is the stellar mass in solar units.
 
We set the mass and radius of all our simulations to those of present-day Earth. We set the orbital parameters (obliquity, eccentricity, and precession) to zero. We use present Earth’s continental configuration and topography, which is a reasonable starting point as fully ocean-covered planets are unlikely to support an active climate-stabilizing carbon-silicate cycle and allow build-up of O$_2$ \citep{AbbotEt2012ApJ,Lingam+Loeb2019AJ}. We place the substellar point stationary over the Pacific at 180$^{\rm o}$ longitude and turn off the quasi-biennial oscillation forcing in all simulations, as this prescription is based on observations of Earth.

We assume initially Earth-like preindustrial surface concentrations of gases N$_2$ (0.78 by volume), O$_2$ (0.21), CH$_4$ ($7.23 \times  10^{-7}$), N$_2$O ($2.73 \times 10^{-9}$), and CO$_2$ ($2.85 \times 10^{-4}$). The existence of an O$_2$-rich atmosphere implies active oxygenating photosynthesis on the surface\footnote{We note that some doubt has been raised as to whether biotic O$_2$ can build-up on planets around M-dwarfs due to the potential paucity of photosynthetically active radiation \citep{LehmerEt2018ApJ,LingamEt2019MNRAS}.} H$_2$O$_v$ and O$_3$ are spatially and temporally variable gases but are initialized at preindustrial Earth values. The surface atmospheric pressure is 101325 Pa (1013.25 mbar). We use the native broadband radiation model of CAM4 and do not include new absorption coefficients as done in \citet{KopparapuEt2017ApJ} due to the extensive effort required to derive new coefficient values for CH$_4$, N$_2$O, and other IR absorbers included in the chemical transport model, which is beyond the scope of this work.

WACCM simulations are run at horizontal resolutions of $1.9\degree \times 2.5\degree$ (latitude by longitude) with 66 vertical levels, model top of $6 \times 10^{-6}$ hPa (150 km), and a model timestep of 900 seconds. We increase the total stellar flux by intervals of $0.1~F_p/F_\oplus$ and modify the rotation period according to Equation 1. We follow previous work \citep{YangEt2014ApJL,KopparapuEt2016ApJ} and assume that the maximum flux for which a planet can maintain thermal equilibrium, i.e., top of atmosphere (TOA) radiation balance, defines the incipient stage of a runaway greenhouse. We refer to climatically-stable (i.e., in thermal equilibrium) simulations as converged simulations, while those that are climatically unstable (i.e., out of thermal equilibrium) are deemed to be in incipient runaway states. With the exception of Figure 1, all converged simulations have been run for 30 Earth years of model time and the results presented here are averaged over the last 10. 

Note that our model simulation naming convention (Table 1) follows from simulated stellar flux and effective temperature: {\it XXFYYT}, where {\it XX} represents the total stellar flux (relative to the Earth’s) and {\it YY} represents the effective temperature of the host star. For instance, 13F26T represents an experiment in which the stellar flux is set to 1.3 $F_\oplus$ and the host star has an effective temperature of 2600 K.

\section{Results}
\label{sec:results}

We present the first simultaneous 3D investigation of climate and atmospheric chemistry in temperate, moist greenhouse, and incipient runaway greenhouse atmospheres on synchronously-rotating planets. The results section is structured as follows: First, we examine the onset of runaway greenhouse conditions in our simulations. Next, we discuss moist greenhouse conditions and their associated climatic and chemical properties. We then investigate isolated changes in stellar spectral type and bolometric stellar flux. Specifically, we analyze the effects of different stellar spectral types, increased incident stellar flux, and UV radiation on our results. Next, we show prognostic water vapor and hydrogen mixing ratios and discuss new escape rates calculated with interactive chemistry. Lastly, we present simulated atmospheric transmission spectra and secondary eclipse thermal emission spectra using our CCM results as inputs and discuss their observational implications.

\begin{figure*}[t] 
\begin{center}
\includegraphics[width=1.8\columnwidth]{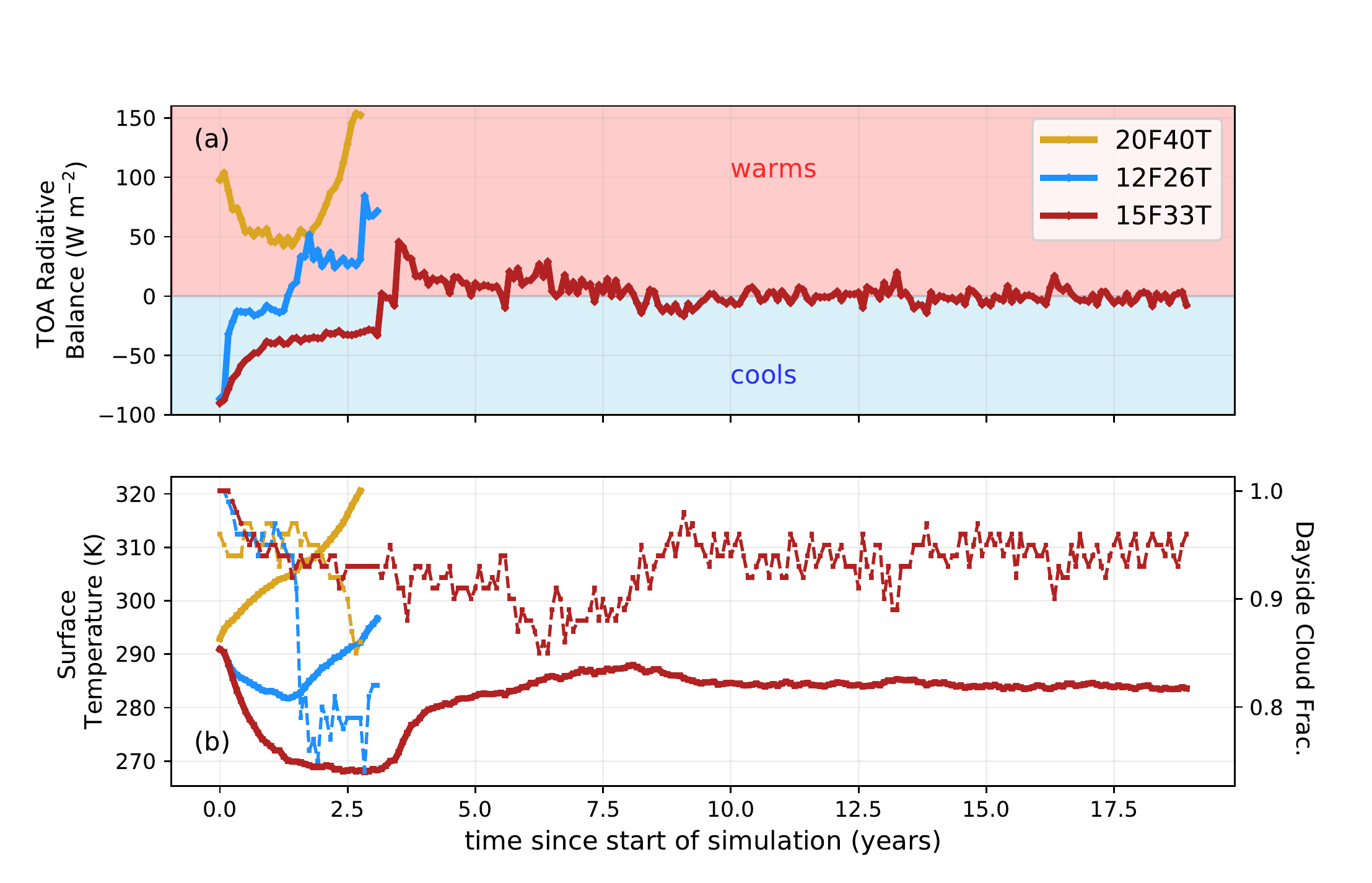}
\caption{\label{fig:eb} Temporal evolution of radiative energy balance (a) and surface temperatures (b) of three representative simulations (12F26T, 20F40T, and 15F33T) across model time of 20 Earth years. Simulations 12F26T and 20F40T progress to incipient runaway greenhouse states due to dissipation of substellar clouds and water vapor greenhouse feedbacks, while 15F33T maintains a stable climate by way of the cloud-stabilizing feedback. Dashed curves represent dayside-mean cloud fraction for the specified simulation.  } 
\end{center}
\end{figure*}

\subsection{Climate Behaviors in Runaway Greenhouse States}

Runaway greenhouse states are energetically unstable climate conditions in which the net absorption of stellar radiation exceeds the ability of water vapor rich and thus thermally opaque atmospheres to emit radiation to space, as described by the Simpson-Nakajima limit \citep{NakajimaEt1992JRG}. A subset of our simulations reaches this runaway threshold, which we define as the innermost boundary of the HZ. We illustrate incipient runaway behavior and contrast it with a climatically-stable case by providing timeseries of TOA radiation balance and surface temperature from three representative simulations, initialized from the same climatic state, i.e., global-mean surface temperature (T$_s$) of ${\sim} 291$ K (Figure~\ref{fig:eb}). From this initial state, simulations diverge, largely according to the intensity of stellar flux and cloud development. For example, in the climatically-stable 15F33T simulation (Figure~\ref{fig:eb}, red curves), the TOA radiation balance begins negative, but stabilizes about zero as the day-side cloud fraction initially decreases, but then oscillates near 95\% coverage. This radiation balance allows the global-mean T$_s$ to stabilize at ~283 K. In contrast to this climatically stable pathway, ``incipient runaway greenhouse" \citep{WolfEt2019ApJ} conditions are simulated for planets at both lower (12F26T) and higher (20F40T) stellar fluxes around late M-dwarf (12F26T) and late K-dwarf (20F40T) stars. In the 20F40T simulation, the planet rapidly transitions into an incipient runaway greenhouse state as the stellar flux is sufficiently high that it causes the collapse of the substellar cloud-albedo shield (Figure~\ref{fig:eb}, gold curves). No equilibrium T$_s$ is achieved in this simulation. Similarly, in the lower flux 12F26T case (Figure~\ref{fig:eb}, blue curves), the global-mean T$_s$ does not achieve an equilibrium. Initially global-mean T$_s$ decreases similar to 15F33T, but the higher rotation rate reduces the dayside cloud shield, allowing the TOA radiation imbalance to turn positive, which leads to the incipient stage of a runaway thermal state

\begin{figure*}[t] 
\begin{center}
\includegraphics[width=1.9\columnwidth]{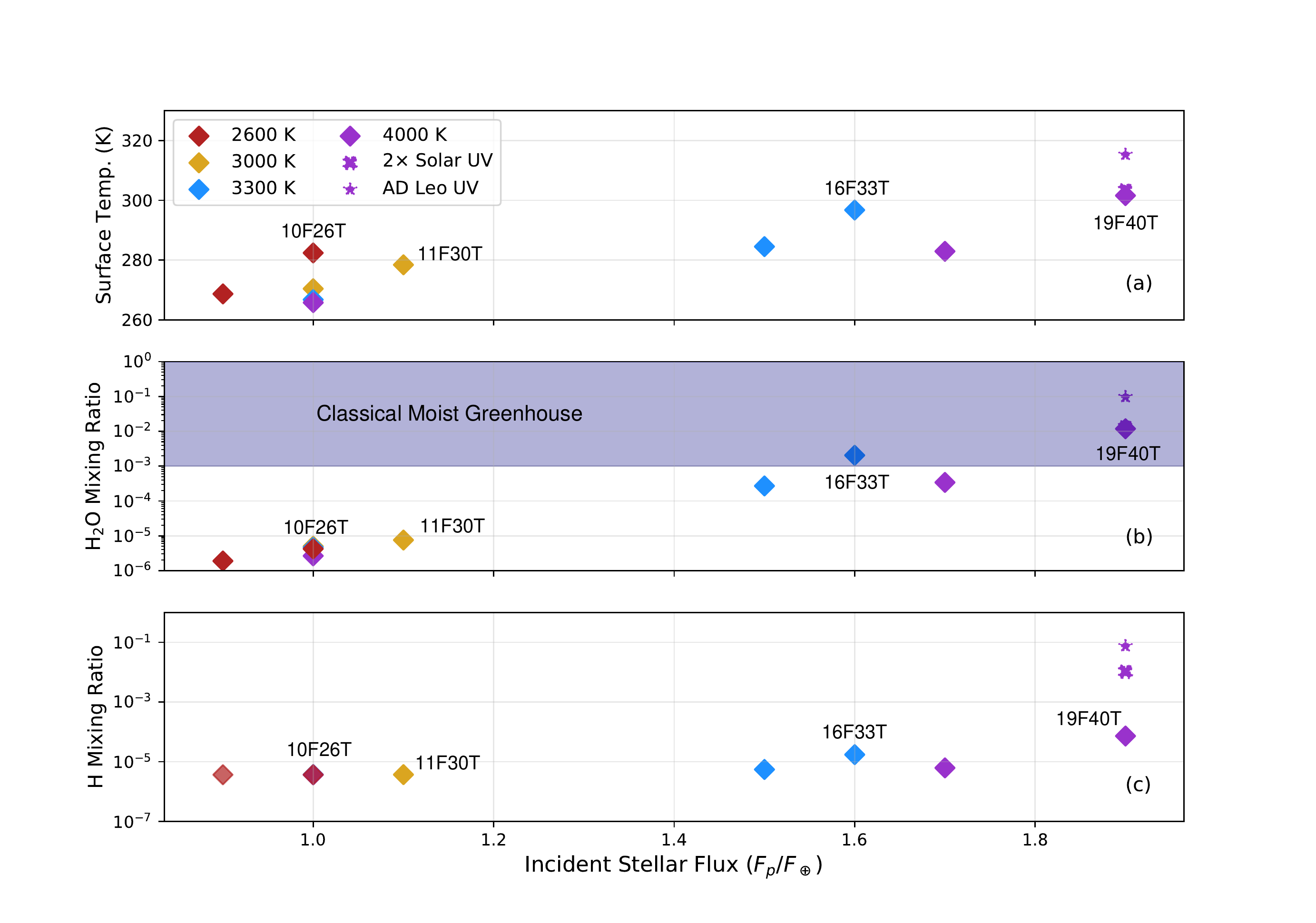}
\caption{\label{fig:summary} Simulated global mean surface temperature (a), stratospheric water vapor mixing ratios (mol mol$^{-1}$) (b), and thermospheric hydrogen mixing ratios (mol mol$^{-1}$) (c) for all climatically-stable experiments plotted according to stellar spectral type and incident stellar flux. IHZ limit simulations are given alphanumeric labels and blue shading in (b) indicates the classical moist greenhouse regime. Color indicates the effective temperature of the host star. Note overlapping results at $F_p= 1.0 F_\oplus$.   } 
\end{center}
\end{figure*}

\subsection{Climate and Chemistry near the IHZ: Temperate \& Moist Greenhouse States}

While runaway greenhouse states delineate the optimistic IHZ, both temperate and moist greenhouse climates may be situated at or near the IHZ limit. In this study, we are primarily interested in the chemistry and climatic conditions habitable planets located at the IHZ. To identify this boundary, four host star type simulations were run with incremental ($0.1~F_p/F_\oplus$) increases in stellar flux (Table 1). Simulations not pushed into the incipient runaway state described in Section 3.1, i.e., simulations with flux $0.1~F_p/F_\oplus$ less than runaway conditions, define our ``IHZ limit" cohort: 10F26T (temperate), 11F30T (temperate), 16F33T (moist greenhouse), and 19F40T (moist greenhouse). Temperate atmospheres have low, Earth-like stratospheric water vapor content (typically $\la 1\e{-5}$ mol mol$^{-1}$) and global-mean T$_s$ below 285 K (Table 1). Moist greenhouse atmospheres emerge when the stratospheric H$_2$O$_v$ mixing ratio are sufficiently high, i.e., $\ga 3\e{-3}$ mol mol$^{-1}$ such that water-loss via diffusion-limited escape could occur at a geologically significant rate in the thermosphere \citep{Kasting1988Icarus}. If the water-loss is sufficiently slow (i.e., $\ga 5$ Gyrs), then rapid desiccation of a planet’s oceans is prevented and its surfaces can remain habitable. 

Amongst our simulations, only IHZ limit climates around early-to-mid M-dwarfs ($T_{\rm eff}  = 3300$ and 4000 K) meet the moist greenhouse criterion (Figure~\ref{fig:summary}b, 16F33T and 19F40T). Simulations around these stars but with lesser stellar flux, i.e., 15F33T and 17F40T, do not achieve sufficient stratospheric H$_2$O$_v$ to place them in the moist greenhouse regime (Figure~\ref{fig:summary}b). Simulations 16F33T and 19F40T have stratospheric H$_2$O$_v$ mixing ratios of $2.05\e{-3}$ and $1.179\e{-2}$ mol mol$^{-1}$ respectively (Table 1), yet their global mean T$_s$ does not exceed 310 K (Figure 2a), which indicates that the surface may be habitable despite the high stratospheric water vapor content. 
Temperate IHZ limit climates (experiments 10F26T and 11F30T) simulated around late M-dwarfs ($T_{\rm eff} = 2600$ and 3000 K) do not enter the moist greenhouse regime with incremental ($0.1~F_p/F_\oplus$) increases in stellar flux (Figure~\ref{fig:summary}, red and gold). Instead, they abruptly transition into incipient runaway greenhouse states (e.g., 12F26T; Figure~\ref{fig:eb}, blue curve). Similar conclusions were reached by \citet{KopparapuEt2017ApJ} using CAM4 with updated H$_2$O$_v$ absorption coefficients but excluding interactive chemistry.

\begin{figure*}[t] 
\begin{center}
\includegraphics[width=2.2\columnwidth]{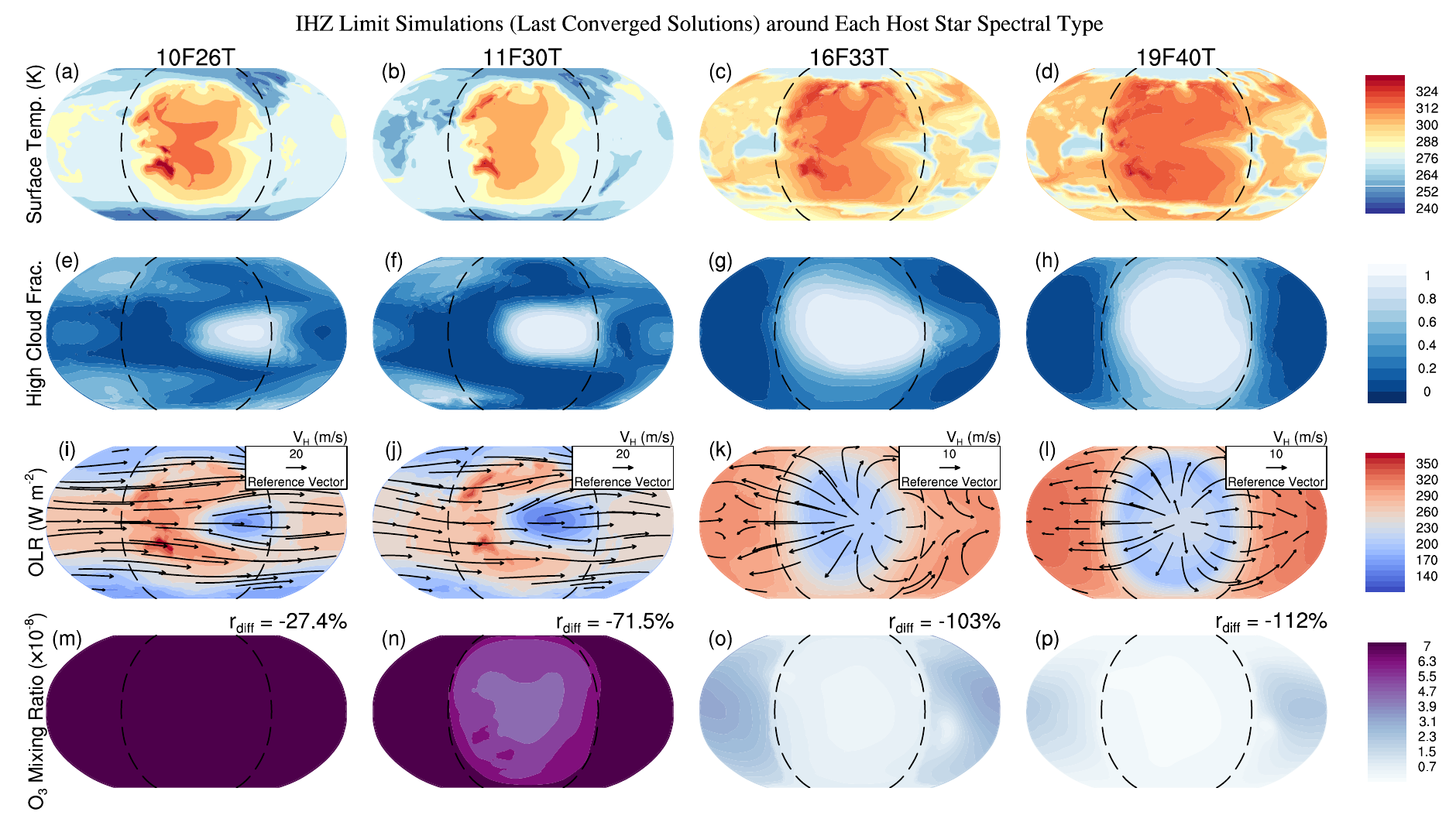}
\caption{\label{fig:map} Simulated global surface temperature (a-d), high cloud fraction (400 to 50 mbar) (e-h), TOA OLR (i-l) and horizontal winds at 100 mbar (i, j) and 10 mbar (k, l), and stratospheric ozone mixing ratios vertically-averaged between $10^{-4}$ and 100 mbar (m-p) from the IHZ limit simulations around each stellar spectral type. $r_{\rm diff}$ is the value of the day-to-nightside mixing ratio contrast defined in Equation~2. Dashed lines indicate the terminators. } 
\end{center}
\end{figure*}

We demonstrate differences in climate and chemistry amongst the IHZ limit simulations across the four host star types by showing contour plots of surface temperature, high cloud fraction, upper atmospheric wind fields, TOA outgoing longwave radiation (OLR), and ozone mixing ratios averaged between $10^{-4}$ and 100 mbar (Figure~\ref{fig:map}). These results exhibit the convolved effects of stellar $T_{\rm eff}$, incident flux, and planetary rotation, as all three parameters are correlated. Following \citet{ChenEt2018ApJL}, we define a metric to assess the day-to-nightside gas mixing ratio contrasts:

\begin{equation}
    r_{\rm diff}  = \frac{r_{\rm day} - r_{\rm night}}{r_{\rm globe} }
\end{equation}

\noindent where $r_{\rm day}$ is the dayside hemispheric mixing ratio mean, $r_{\rm night}$ the nightside mean, and $r_{\rm globe}$ the global mean. The degree of anisotropy is loosely encapsulated in this parameter, which is shown in Figures~\ref{fig:map} and \ref{fig:map_uv} and will be discussed throughout the paper.

Substantial differences in surface temperature distributions can be found amongst the four IHZ limit simulations. With increasing stellar $T_{\rm eff}$, day-to-nightside T$_s$ gradients decrease (Figures~\ref{fig:map}a-d). This is caused by increased day-to-nightside heat redistribution at higher incident fluxes. Consistent with previous GCM studies that exclude interactive chemistry (e.g., \citealt{KopparapuEt2017ApJ}), we find that on slowly rotating planets, the weaker Coriolis force allows formation of optically thick substellar cloud decks by way of buoyant updrafts (Figures~\ref{fig:map}e-h). In addition, meridional overturning cells expand to higher latitudes when the Rossby radius of deformation approaches the diameter of the planet \citep{DelGenioEt1993Icar}, which decrease the pole-to-equator temperature gradient. These two consequences (i.e., formation of dayside clouds and reduction of temperature gradient) of slow rotation allow planets around early M-dwarfs to maintain habitable climates at higher fluxes.

Atmospheric dynamics regulate cloud patterns, circulation symmetry, and transport of airmasses on a planet. For slowly-rotating cases, substellar OLR is reduced by the high opacity of deep convective cloud decks, which induce a strong warming effect (experiments 16F33T and 19F40T; Figure~\ref{fig:map}k-l). As we move from 16F13T, to 11F30T, then to 10F26T, the dynamical state gradually transitions from divergent circulation to one dominated by tropical Rossby waves and zonal jets. The resultant elevated high-to-low latitude momentum transport can be seen in the streamlines and OLR patterns (Figure~\ref{fig:map}i-j; see also \citealt{Matsuno1966,Gill1980}), and is likely caused by shear between Kelvin and Rossby waves \citep{Showman+Polvani2011ApJ}. Comparing our results to the circulation regimes studied by \citet{Haqq-MisraEt2018ApJ}, we find that simulations 16F33T and 19F40T are situated in their slow rotating regime (Figure~\ref{fig:map}i-j), in which thermally driven radial flows dominate. Simulation 10F26T (Figure~\ref{fig:map}i) is consistent with the rapid rotator characterized by strong zonal jet streams and a weaker substellar rising motion, while 11F30 belongs in the so-called Rhines rotator regime, in which the OLR and radial flows are shifted eastward by the emergence of turbulence (Figure~\ref{fig:map}j). For the simulation in latter Rhines rotator regime, the meridional extent of Rossby waves is just under the planetary radius value, thus horizontal flow is a combination of superrotation and thermal-driven circulation (similar in terms of dynamical behavior to the ``transition regime" found by \citealt{CaroneEt2015MNRAS}).

\begin{figure*}[t] 
\begin{center}
\includegraphics[width=2.1\columnwidth]{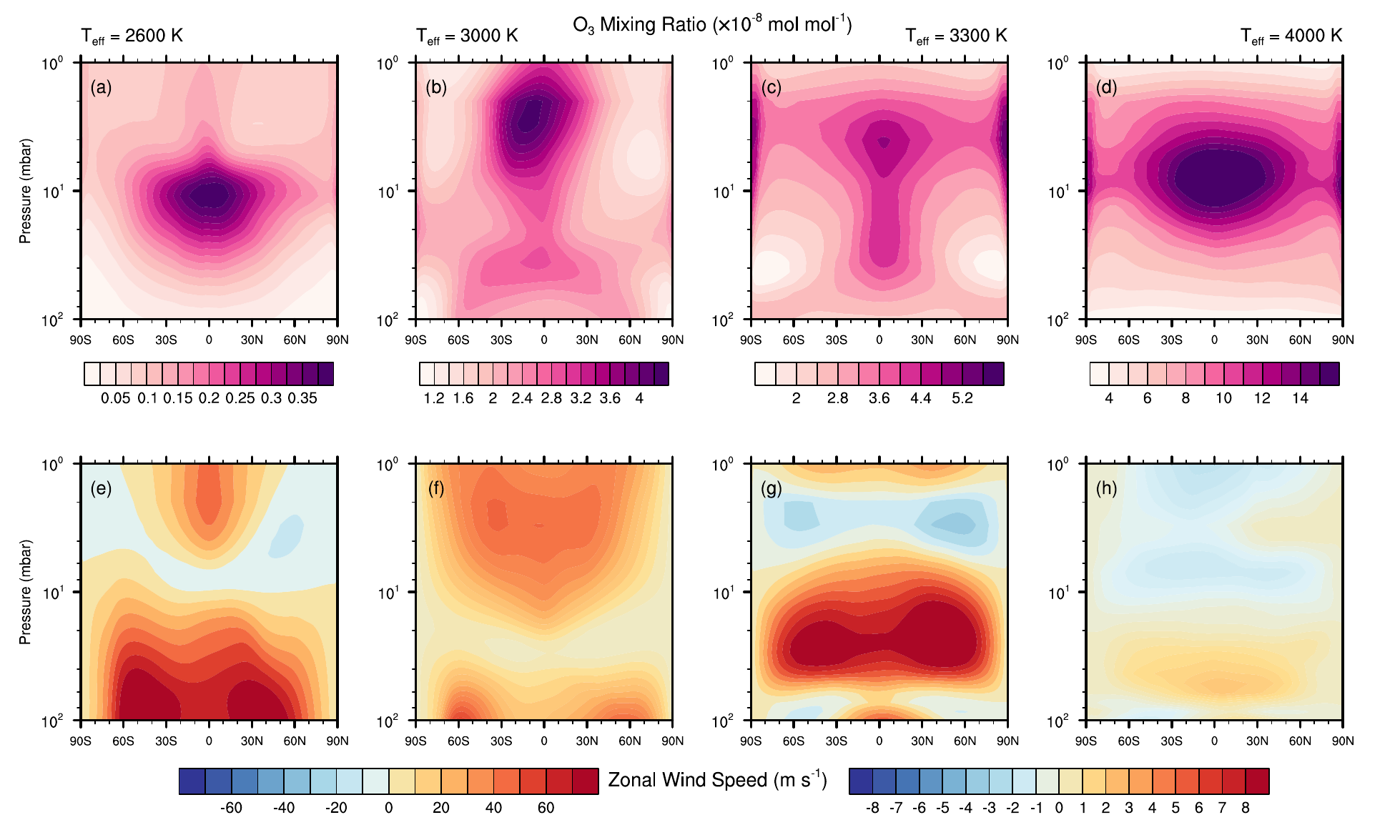}
\caption{\label{fig:oz_trans} Zonal mean of ozone mixing ratio around stars with stellar effective temperatures of 2600 K (a), 3000 K (b), 3300 K (d) and zonal mean of zonal wind around the same set of stellar $T_{\rm eff}$s (e, f, g, h). Direct day-night side circulation allows global dispersal of ozone (c, d), whereas strong zonal jets disrupt efficient equator-to-pole zone transport (a, b). Note the different color bar ranges.}
\end{center}
\end{figure*}

An additional consideration, allowed by the coupling of chemistry and dynamics, is the role of stratospheric circulation in the transportation of airmasses (and thus photochemically produced species and aerosols). Atmospheres around late M-dwarfs (simulations 10F26T and 11F30T) display superrotation that induces standing tropical Rossby waves, thereby confining the majority of the produced ozone near the equator (Figure~\ref{fig:oz_trans}a and b). \citet{CaroneEt2018MNRAS} explained this by the weakening of the extratropical Rossby wave and reduced efficiency of stratospheric wave breaking; here we confirm their hypothesis by directly accounting for ozone photochemistry and transport. In atmospheres with high circulation symmetry (simulations 16F33T and 19F40T), tropical jets are effectively damped. This leads to increased strength of stratospheric meridional overturning circulations (i.e., a thermally driven version of the Brewer-Dobson circulation) and that of the Walker circulation, which allows equator-to-pole and day-to-nightside dispersal of ozone (Figure~\ref{fig:oz_trans}c and d). Lastly, ozone day-to-night mixing ratio contrasts ($r_{\rm diff}$) depend on both chemical (e.g., reaction with OH) and dynamical (e.g., strength of Rossby and Kelvin waves) factors and highlight the interplay between transport, photochemical, and photolytic processes (Figure~\ref{fig:map}m-p; see also \citealt{ChenEt2018ApJL}).


\subsection{Temperate Atmospheres: Effects of Changes in Stellar SED} 

Host star spectral-type can influence attendant planet atmospheres through changes in stellar $T_{\rm eff}$ and planetary rotation period, as the latter two variables are correlated through Kepler's third law. Our coupled CCM simulations demonstrate that the primary climatic effects of different input stellar SEDs are modulations in greenhouse gas radiative forcing, while photochemistry (e.g., driver of water photolysis) is not substantially impacted.

\begin{figure*}[t] 
\begin{center}
\includegraphics[width=2.1\columnwidth]{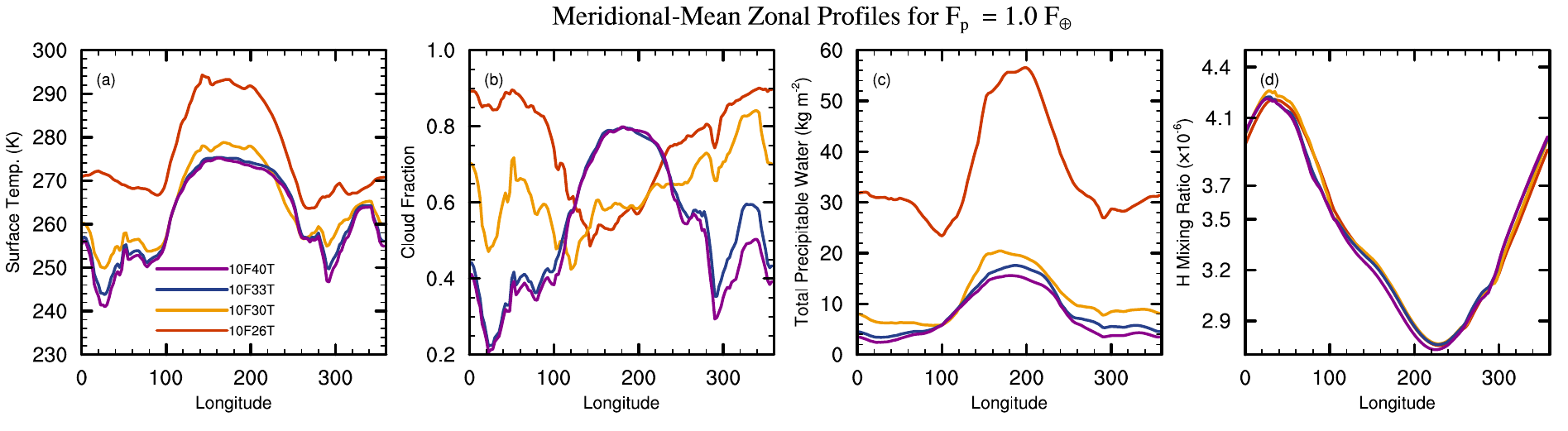}
\includegraphics[width=2.1\columnwidth]{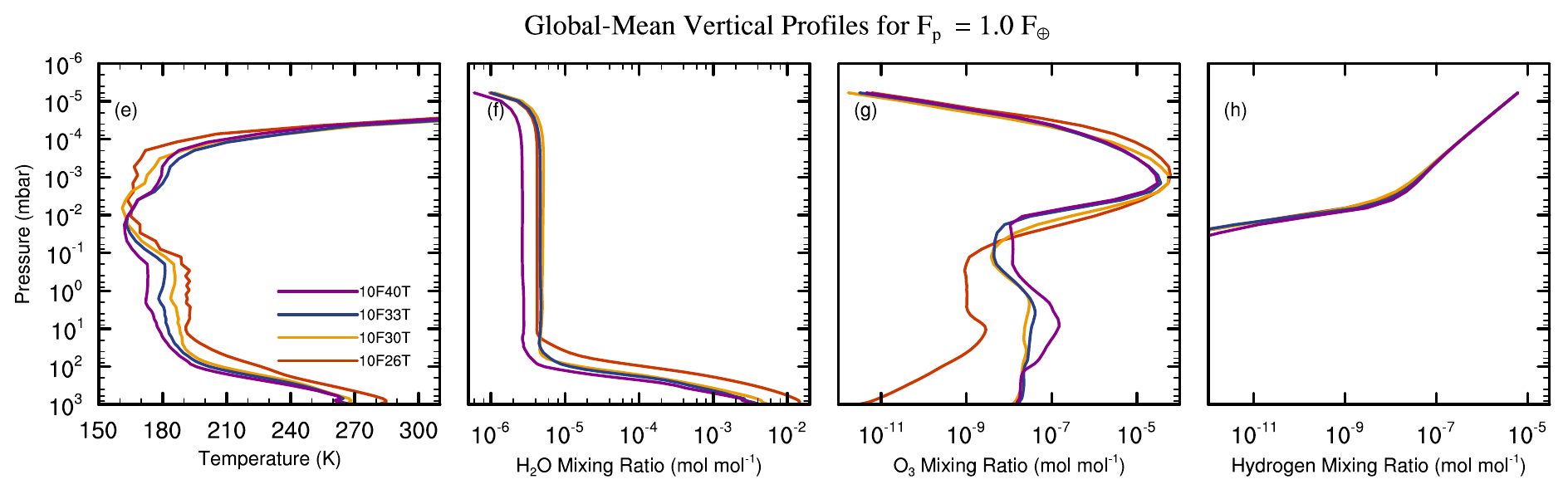}
\caption{\label{fig:sed} Zonal profiles of mean meridional surface temperature (a), total cloud fraction (b), vertically-integrated precipitable water (c), model-top H mixing ratios (mol mol$^{-1}$) (d), and global-mean vertical profiles of atmospheric temperature (e), H$_2$O$_v$ (f), O$_3$ (g), and H mixing ratios (mol mol$^{-1}$) (h). We show simulations around stars with $T_{\rm eff} = $2600, 3000, 3300, 4000 K but with the same total stellar flux $F_p = 1.0 F_\oplus$. In the zonal profiles, the substellar point is over longitude 180$\degree$. } 
\end{center}
\end{figure*}

The red-shifted spectra of low-mass stars have consequences on climate and chemistry by way of an increased water vapor greenhouse effect and reduction in dayside cloud cover (Figure~\ref{fig:sed}a-c). Specifically, greater IR absorption by atmospheres around stars with $T_{\rm eff} = 2600$ K and 3000 K increases both atmospheric temperature and the amount of precipitable water, and decrease radiative cooling efficiency aloft (Figure~\ref{fig:sed}a and b, red curve). Reduction in the efficiency of radiative cooling and dayside cloud fractions lead to the water vapor greenhouse effect offsetting that of cloud albedo, and results in higher T$_s$ for 10F26T compared to simulations around early M-dwarfs (Figure~\ref{fig:sed}b). Moreover, potential increased concentrations of other greenhouses gases such as CH$_4$ and N$_2$O on late M-dwarf planets would also contribute to increased T$_s$ \citep{SeguraEt2005AsBio,RugheimerEt2015ApJ}. 

Further insight into the effects of different stellar $T_{\rm eff}$ can be observed in the global-mean vertical profiles (Figure~\ref{fig:sed}e-h). Below 80 km ($10^{-2}$ mbar), atmospheric temperature increases monotonically with decreasing stellar $T_{\rm eff}$ (Figure~\ref{fig:sed}e). This relationship exists because longwave absorption increases and Rayleigh scattering decreases with the redness of the host star. Above 80 km ($10^{-2}$ mbar), however, the dependence of temperature on stellar $T_{\rm eff}$ is reversed due to the increasingly important role of O$_2$ photodissociation by shortwave photons at higher altitudes (Figure~\ref{fig:sed}e). Both H$_2$O$_v$ shortwave heating and strength of vertical advection increase with lower $T_{\rm eff}$ due to the higher NIR fluxes. At pressures less than $10^{-4} - 10^{-5}$ mbar, temperatures rise rapidly (Figure~\ref{fig:sed}e) by way of thermospheric O$_2$ and O absorption of soft X-ray and EUV, while water vapor mixing ratios decline due to photodissociation to H, H$_2$, and OH (Figure~\ref{fig:sed}f).

Ozone photochemistry is modulated by incident UV flux, chemical reaction pathways, and ambient meteorological conditions ($P$, $T$). Our predicted ozone mixing ratios above 70 km ($10^{-1}$ mbar) reduce with decreasing stellar $T_{\rm eff}$ (Figure~\ref{fig:sed}g). Elevated OH production through water vapor photosys leads to greater photochemical removal of O$_3$ by OH. OH destruction of ozone is maximized near the boundary layer for the 10F26T experiment as increased surface temperatures (due to lower substellar albedos) leads to larger H$_2$O$_v$ inventories. At pressures less than $10^{-2}$ mbar, ozone mixing ratios rise as mean free paths between molecules increase dramatically with altitude.

Atmospheric hydrogen is primarily produced from the photolysis of high-altitude water vapor. At temperate conditions, H mixing ratios in both meridional and vertical profiles (Figure~\ref{fig:sed}d and Figure~\ref{fig:sed}h) are not substantially impacted by shifts in stellar SED, assuming quiescent stars. This is seen by the fact that all four simulations have close to zero H mixing ratios until ${\sim}10^{-3}$ mbar, at which point they rise to ${\sim} 10^{-7}$ mol mol$^{-1}$ (Figure~\ref{fig:sed}h). Increased efficiency in water vapor photolysis above the mesosphere (${\sim}10^{-2}$ mbar) is evidenced by the exponential dependence of H mixing ratios on altitude. A transition in H mixing ratios above 80 km ($10^{-2}$ mbar) is caused by the rapid increase in water vapor photolysis rates and hence stronger dependence on pressure altitude. We should point out however, that the seemingly minor effects of stellar $T_{\rm eff}$ stem from our choice of input SED-types. The PHOENIX stellar model data \citep{HusserEt2013A&A} we employed are inactive in the UV and EUV bands, regardless of the spectral type. As many M-dwarfs are active in the Ly-$\alpha$ line fluxes ($115 < \lambda < 310$ nm; \citealt{FranceEt2013ApJ}), we next explore how changes in these assumptions affect our findings (Section 3.5).

\begin{figure*}[t] 
\begin{center}
\includegraphics[width=2.1\columnwidth]{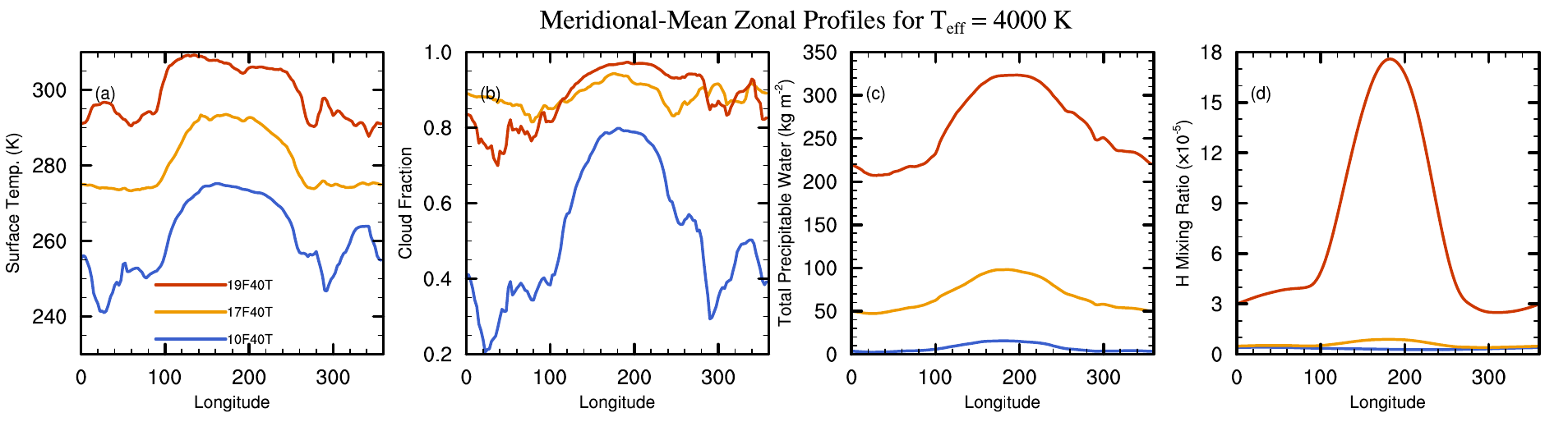}
\includegraphics[width=2.1\columnwidth]{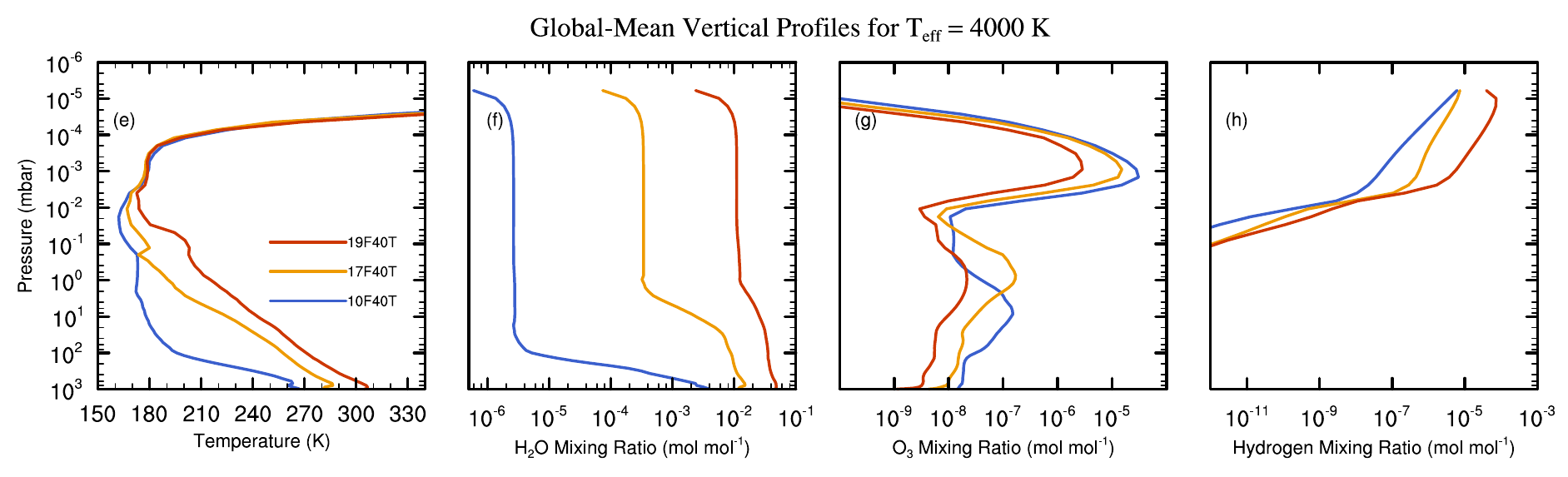}
\caption{\label{fig:flux} Zonal profiles of mean meridional surface temperature (a), total cloud fraction (b), vertically-integrated precipitable water (c), model-top H mixing ratios (mol mol$^{-1}$) (d), and global-mean vertical profiles of atmospheric temperature (e), H$_2$O$_v$ (f), O$_3$ (g), and H mixing ratios (h). Simulations use total stellar fluxes $F_p = 1.0, 1.5, 1.8 F_\oplus$ and stellar $T_{\rm eff}$ held fixed at 4000 K. In the zonal profiles, the substellar point is over longitude 180$\degree$.} 
\end{center}
\end{figure*}

\subsection{Moist Greenhouse Atmospheres: Effects of Increasing Stellar Flux}

Increasing stellar flux can affect both climatic and photochemical variables. For instance, we find that both atmospheric temperature and water vapor mixing ratios increase monotonically with increasing incident flux as reported by previous studies (e.g., \citealt{KastingEt1984Icarus,KastingEt2015ApJL}), whereas photochemically important species and their derivatives such as ozone and hydrogen display nonlinear behavior.

Water vapor concentrations and surface climate are both strong functions of stellar flux. With increasing stellar flux at fixed stellar $T_{\rm eff}$ (= 4000 K) we find that water vapor quickly becomes a major constituent from the stratosphere (${\sim}100$ mbar) to the thermosphere ($5\e{-5}$ mbar). For example, total precipitable water increases by a factor of 5 for every interval change of incident flux (Figure~\ref{fig:flux}c). However, the corresponding surface temperature rises much more gradually, wherein a change in stellar flux  (from experiment 10F40T to 17F40T, or from $F_p = 1.0~F_\oplus$ to $F_p = 1.6~F_\oplus$) only causes an average T$_s$ increase of ${\sim}20$ K in the substellar hemisphere due to stabilizing cloud feedbacks (Figure~\ref{fig:flux}a).

The rapid rise in water vapor mixing ratios in the upper atmosphere ($\ga 1$ mbar) can be attributed to positive feedbacks between flux, H$_2$O$_v$ IR heating, and vertical motion. H$_2$O$_v$ NIR absorption and cloud feedbacks are amplified by increased stellar flux and incident IR. With increased water vapor, the tropospheric lapse rate decreases and the moist convection zone expands. These two shifts lead to displacement of the cold trap to higher altitudes (Figure~\ref{fig:flux}e) and thus more efficient H$_2$O$_v$ vertical advection. Greater H$_2$O$_v$ vertical transport increases its concentration in the upper atmosphere, increasing the strength of greenhouse effect and atmospheric temperature.

While we find considerable increases in H$_2$O$_v$ mixing ratios in the stratosphere, the increase in surface temperatures is less dramatic. For instance, in experiment 19F40T, the stratospheric water vapor mixing ratio has reached $10^{-2}$ mol mol$^{-1}$ (Figure~\ref{fig:flux}f) yet the global-mean surface temperature is still just 300 K (Figure~\ref{fig:flux}e). This result agrees with previous findings of the so-called habitable moist greenhouse in which the stratosphere becomes highly saturated while the troposphere is stabilized by optically thick clouds \citep{KopparapuEt2017ApJ}. In habitable moist greenhouse states (e.g., 19F40T; Figure~\ref{fig:flux}e-h), surface habitability, to the first order, depends on the rate of water escape from the upper atmosphere. If water escape is sufficiently slow in these conditions, then insofar as the surface climate remains stable, the planet could host life on a timescale that raises the possibility of remote detection.

Stellar flux can indirectly affect ozone photochemistry via an increase in atmospheric water vapor dissociation and changes in ambient conditions such as atmospheric temperature. Vertical profiles of ozone show a minimum in the ozone mixing ratio in the highest total incident flux simulation (Figure~\ref{fig:flux}g, 19F40T, red curve), and a maximum for the simulations receiving the least (Figure~\ref{fig:flux}g, 10F40T, blue curve). The resultant thinner ozone layer at high fluxes is caused by the increased removal rate via photochemical reactions with dayside HO$_{\rm x}$ and NO$_{\rm x}$ (primarily OH and NO species) for simulation 19F40T. Reduction in ozone between the boundary layer and altitude at 5.0 mbar is due to photochemical removal by OH, while the ozone maximum is shifted to 1.0 mbar (Figure~\ref{fig:flux}g, gold curve), indicating a change in the location of highest gross ozone production rate.

Elevated water vapor mixing ratios lead to more atomic hydrogen via photodissociation. At temperate conditions, the most efficient altitude of water vapor photolysis is at pressure levels less than $10^{-3}$ mbar, as implied by the H mixing ratio shift (Figure~\ref{fig:flux}h). With higher incident fluxes however, e.g., 19F40T, the inflection of H mixing ratios change with altitude indicating higher photodissociation efficiencies with height. Notably, we find that our prognostic H mixing ratios are almost never twice the amount of H$_2$O$_v$, as assumed in previous studies (e.g., \citealt{KastingEt1993Icarus,KopparapuEt2013ApJ,KopparapuEt2017ApJ}). This suggests that previous climate modeling works on the moist greenhouse state have overestimated water-loss rates.  Lesser simulated H mixing ratios are the result of a variety of processes, including the oxidation of H by O$_2$ and photochemical shielding by O$_3$, CH$_4$, and N$_2$O. In the next section we explore the dependence of photochemistry on stellar UV radiation inputs. With simulated hydrogen, we then provide revised calculations of water loss and estimate the longevity of our exoplanetary oceans.

\begin{figure*}[t] 
\begin{center}
\includegraphics[width=2.0\columnwidth]{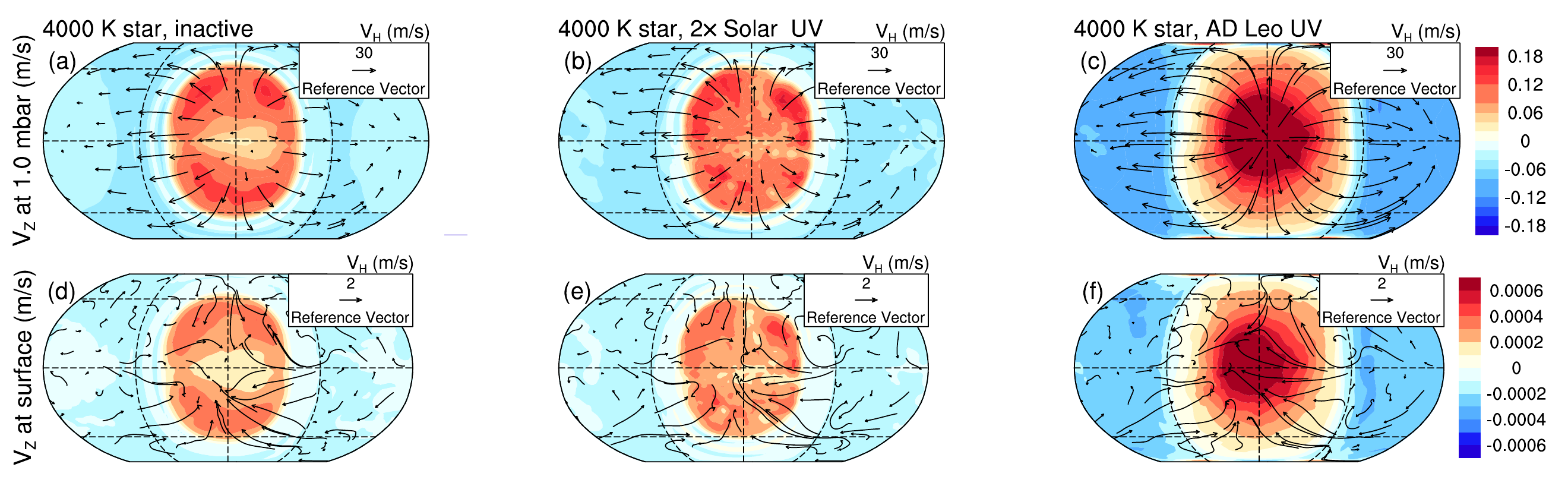}
\caption{\label{fig:wind_uv} Simulated global wind velocity fields at 1.0 mbar (a-c) and at the surface (d-f) at three different stellar UV radiation levels. Colored contours present simulated vertical velocities (in the z-direction) at the indicated height while vectors represent the horizontal wind velocities. } 
\end{center}
\end{figure*}  

\hfill \break

\subsection{Dependence on Stellar UV Activity}

Stellar UV radiation can affect atmospheric chemistry, photochemistry, surface habitability, and based on our findings, observability. Here, we investigate the 3D effects of different stellar UV activity assumptions on tidally-locked planets with Earth-like atmospheres. To test the effects of UV radiation we run simulations in which the UV bands ($\lambda < 300$ nm) of the fiducial $T_{\rm eff} = 4000$ K (experiment 19F40T) star are swapped with those of (a) active VPL AD Leonis data (19FADLeoUV; \citealt{SeguraEt2005AsBio}) and (b) UV data obtained by doubling the Solar UV spectrum (19FSolarUV; \citealt{lean1995reconstruction}). 
Active stellar SEDs are joined with the VIS/NIR portion of the spectra beyond 300 nm by linearly merging the last UV datapoint with the first optical ($\lambda > 300$ nm) datapoint in the PHOENIX stellar model. The aim in this section is to test how including UV activity could alter the conclusions in Section 3.1  Only changes in the UV wavelengths of the SEDs are tested as we are primarily interested in the isolated effects of UV photons, rather than those in other wavelengths. Follow-up work will make use of HST + XMM/Chandra-based M dwarf spectra with observed UV bands from \citet{FranceEt2013ApJ}, \citet{YoungbloodEt2016ApJ}, and \citet{LoydEt2016ApJ}.

UV radiation may drive changes in atmospheric dynamics, in addition to atmospheric chemistry. With changes in our fiducial late K-dwarf ($T_{\rm eff} = 4000$ K) SED, we find increases in the vertical velocities at the substellar point from 0.10 m s$^{-1}$ (inactive star;), to 0.14 m s$^{-1}$ ($2\times$ Solar UV), then to 0.18 m s$^{-1}$ (AD Leo UV; Figure~\ref{fig:wind_uv}a-c), indicating stronger ascent of substellar updrafts. In addition, horizontal and thus day-to-nightside transport are enhanced as evidenced by the higher wind velocities. While mesospheric (1 mbar) zonally-averaged wind speeds forced by the quiescent M-dwarf are modest ${\sim}$25 m s$^{-1}$ (Figure~\ref{fig:wind_uv}a), equatorial winds driven by elevated day-to-nightside temperature gradients can reach as high as 60 m s$^{-1}$ for the simulations forced by the AD Leo UV SED (Figure~\ref{fig:wind_uv}c). Near-surface winds converge toward the substellar point due to large-scale updrafts in all three cases, but are not significantly altered by changes in UV radiation (Figure~\ref{fig:wind_uv}d-f).

\begin{figure*}[t] 
\begin{center}
\includegraphics[width=2.0\columnwidth]{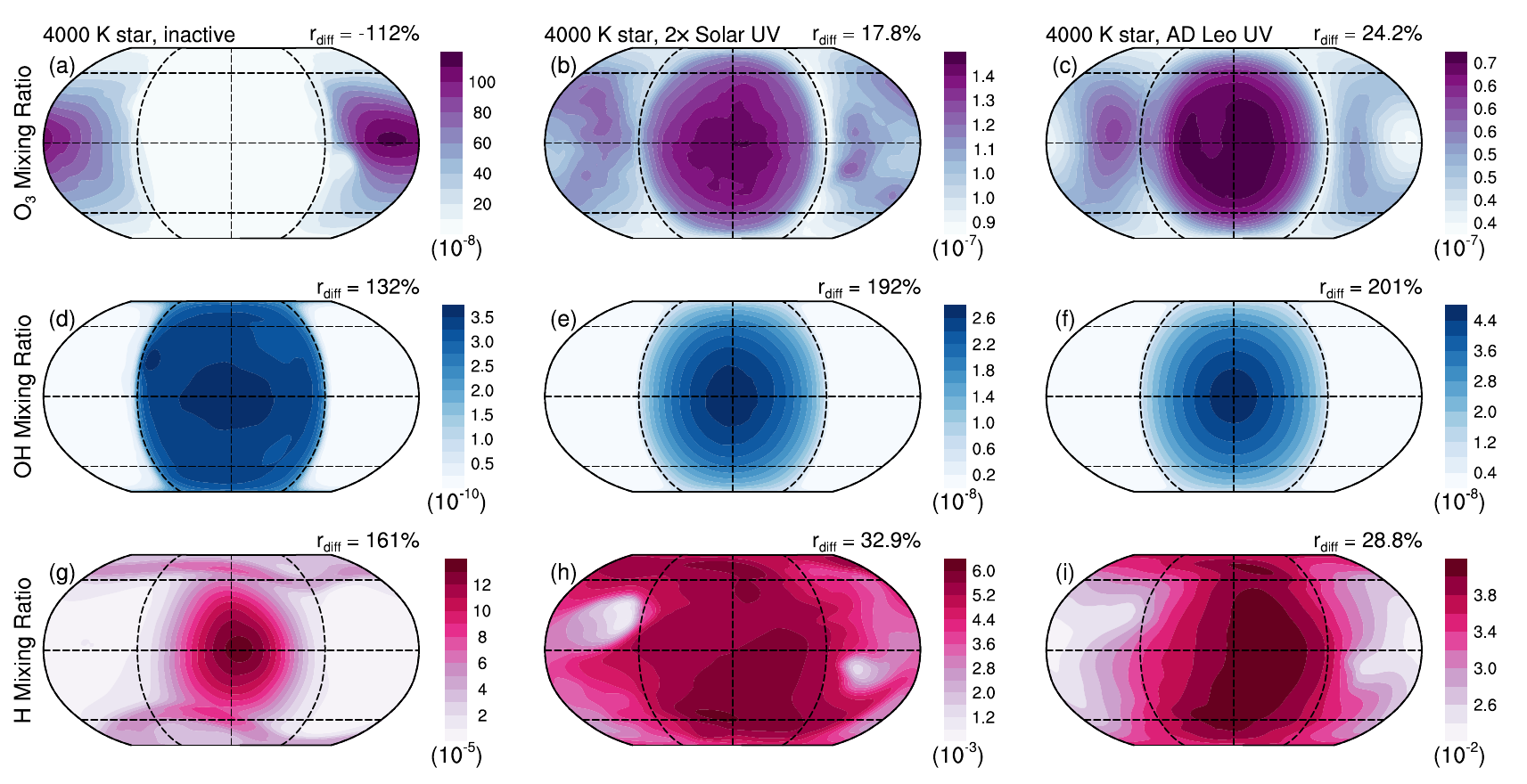}
\caption{\label{fig:map_uv} Simulated global ozone mixing ratio (a), OH mixing ratio (b), and H mixing ratio (mol mol$^{-1}$) (c) at three different levels of UV radiation: inactive, $2 \times$ Solar, and AD Leo. Ozone and OH mixing ratios are vertically averaged (column-weighted) between $10^{-4}$ and 100 mbar. H mixing ratio values are reported at model-top (${\sim} 5 \times 10^{-6}$ mbar). Note that each panel has a unique color bar range. } 
\end{center}
\end{figure*}

Global distributions of photochemically important species and their byproducts are also affected by stellar UV activity (Figure~\ref{fig:map_uv}). Substellar updrafts (due to radiative heating) of chemical constituents and antistellar downdrafts (due to radiative cooling) should result in higher ozone mixing ratios on the dayside. For the simulations forced by the $2\times$ Solar UV and AD Leo UV SED (Figure~\ref{fig:map_uv}b-c), this effect is heightened by increased UV in the wavelengths responsible for ozone production (shortward of 220 nm). In contrast, lower ozone production rates on the quiescent simulation lead to reduced dayside ozone (Figure~\ref{fig:map_uv}a).

The amount of dayside OH and H is directly related to the input UV, with chemical transport playing a small role. With a higher UV radiation than that received by the baseline, the OH distributions become increasingly concentric (19FSolarUV: 192\% and 19FADLeoUV: 201\%; Figure~\ref{fig:map_uv}e-f) due to its short lifetime and the greater contribution from water vapor photodissociation. In contrast at higher UV levels, H mixing ratio distributions begin to lose their concentric shapes and reduced $r_{\rm diff}$ (19FSolarUV: 32.9\%, and 19FADLeoUV: 28.8\%; Figure~\ref{fig:map_uv}g-i). These different responses are explained by the enhanced dispersal of H by atmospheric transport as seen by the slight eastward shift of H mixing ratio distribution in the AD Leo UV case (Figure~\ref{fig:map_uv}i). Increased horizontal advection migrates the effects of enhanced dayside photolytic removal, reflected in the decreasing $r_{\rm diff}$ of hydrogen with greater UV input. Note that our inactive stellar SEDs result in more reduced dayside ozone than those reported by \citet{ChenEt2018ApJL}, which likely stems from the lack of a fully resolved stratosphere-MLT region (e.g., Brewer-Dobson circulation) in the low-top out-of-the-box version of CAM4. These discrepancies in day-to-nightside chemical gradients illustrate the need for model inter-comparisons of exoplanetary climate predictions (e.g., \citealt{YangEt2019ApJb}).

\begin{figure*}[t] 
\begin{center}
\includegraphics[width=2.0\columnwidth]{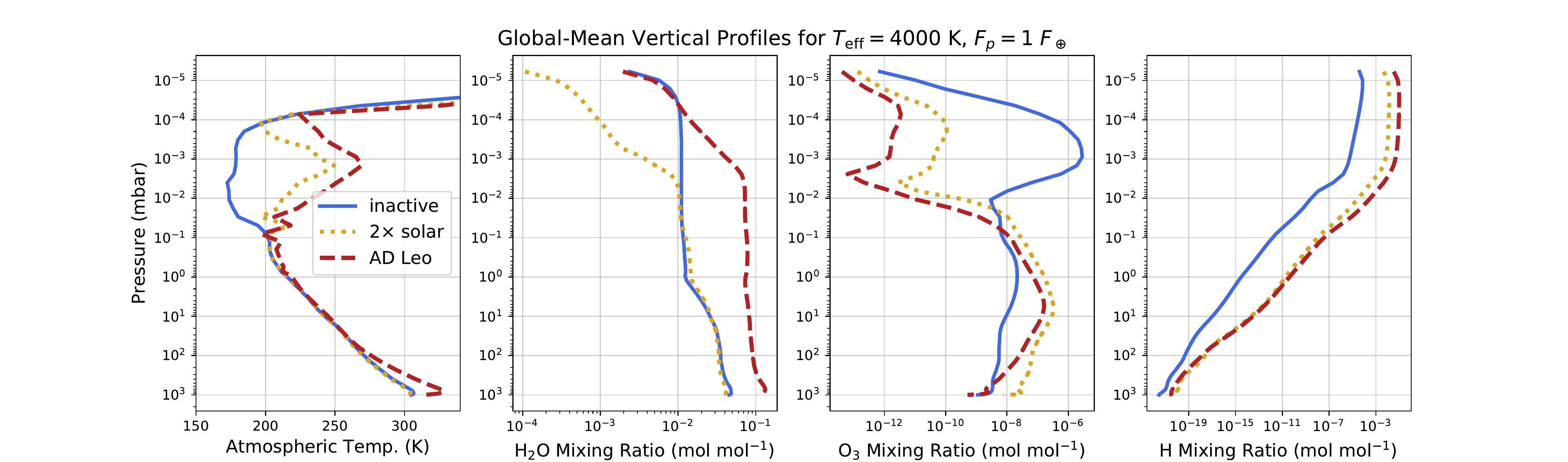}
\caption{\label{fig:vert_uv} Global-mean vertical profiles of atmospheric temperature (a), H$_2$O$_v$ (b), O$_3$ (c), and H mixing ratios (mol mol$^{-1}$) (d) at three different levels of UV radiation: inactive, $2 \times$ Solar, and AD Leo. }
\end{center}
\end{figure*}

Unsurprisingly, the three different UV radiation schemes produce atmospheric temperature profiles that are substantially different (Figure~\ref{fig:vert_uv}a). Elevated incident UV fluxes translate to higher shortwave heating and thus atmospheric temperatures due to FUV and EUV absorption by atomic and molecular oxygen. A ``harder" UV spectrum is also able to penetrate more deeply into the atmosphere.

Apart from different thermal structures, we find orders of magnitude differences in H$_2$O$_v$, O$_3$, and H mixing ratio profiles (Figure~\ref{fig:vert_uv}b, c, and d)$-$ indicating that stellar activity can have strong ramifications for water loss and atmospheric chemistry for moist greenhouse atmospheres. Enhanced AD Leo EUV and UV induced shortwave heating increases stratospheric and mesospheric (between 100 and 1 mbar)  temperatures and water vapor mixing ratios (red curve; Figure~\ref{fig:vert_uv}a and b). However, the altitude at which photolysis is maximized moves lower due to the more energetic shortwave photons (Figure~\ref{fig:vert_uv}b). For ozone mixing ratios, production outpaces destruction resulting in a thicker ozone layer for simulations around more active stars (gold and red curves, Figure~\ref{fig:vert_uv}c), while the upper ozone layers remain desiccated.

Finally, different input UV assumptions also alter the altitude at which water vapor photolysis is most efficient, which can determine the thermospheric H mixing ratio and hence water escape rate. Without stellar activity, the H mixing ratios remain low $7.27\e{-5}$ mol mol$^{-1}$ (red curve, Figure~\ref{fig:vert_uv}d). With the inclusion of stellar activity, both altered UV simulations are pushed into the true moist greenhouse regime with H mixing ratios of $1.04\e{-3}$ and $7.45\e{-2}$ mol mol$^{-1}$ respectively (gold and red curves, Figure~\ref{fig:vert_uv}d). This implies that although planets around inactive stars may only experience minor water loss, both active Solar and AD Leo SEDs could cause attendant planets to suffer rapid water loss.

Inclusion of stellar UV activity may modify conclusions regarding host star spectral type dependent IHZ boundaries, as moist greenhouse atmospheres around early M-dwarfs ($T_{\rm eff} \sim 4000 $K) are more vulnerable to photodissociation than temperate climates around late M-dwarfs ($T_{\rm eff} \sim 2600$ K).  With the present simulations, it is challenging to further this possibility as our grid of stellar effective temperature values are rather coarse (i.e., only four $T_{\rm eff}$s between 2600 and 4000 K).

\begin{figure}[h] 
\begin{center}
\includegraphics[width=1.1\columnwidth]{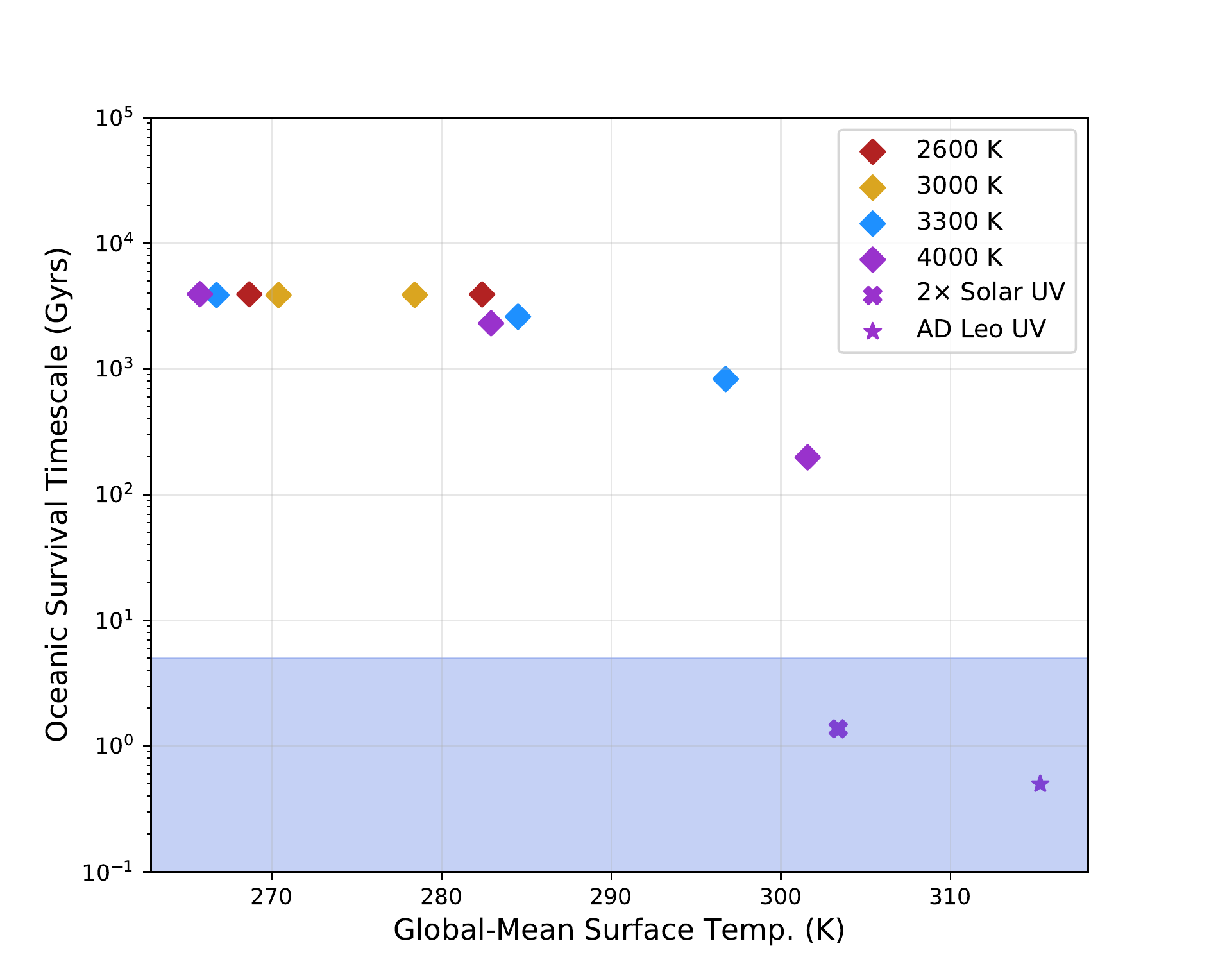}
\caption{\label{fig:survival} Ocean survival timescale as a function of stellar $T_{\rm eff}$ (2600, 3000, 3300, 4000 K), UV activity ($2 \times$ Solar and AD Leo), and their corresponding global-mean T$_s$. Our results suggest that only simulations around active M-dwarfs enter the classical moist greenhouse regime as defined by \citet{KastingEt1993Icarus}. Blue shading indicates timescales less than the age of the Earth.} 
\end{center}
\end{figure}  

\subsection{Prognostic Hydrogen \& Ocean Survival Timescales}

The ability of a given planet to host a viable habitat is linked to the survivability of its ocean, i.e., the so-called ``ocean loss timescale". Conventional estimates of the ocean loss timescale have used 1D climate models and GCMs that rely on prescribed H mixing ratios calculated by doubling their model-top H$_2$O$_v$. Here we reassess previous estimates by using directly simulated H mixing ratios and thermospheric temperature profiles drawn from our CCM simulations. We find that our ocean survival timescales are substantially higher than previously published estimates for quiescent stars, and are critically dependent on the stellar activity level. We demonstrate this by estimating water loss rates with Jeans' diffusion-limited escape scheme. While an over-simplification due neglect of hydrodynamics, this first order estimate is typically used to interpret climate model results of moist greenhouse atmospheres (e.g., \citealt{KastingEt1993Icarus,KopparapuEt2013ApJ,KopparapuEt2017ApJ,Wolf+Toon2015JGR}). With our prognostic hydrogen mixing ratios at each stellar $T_{\rm eff}$ and incident flux combination (Figure 2), we can calculate new escape rates of hydrogen \citep{Hunten1973JAS}:

\begin{equation}
    \Phi(H) \approx \frac{b Q_{\rm H}}{H} 
\end{equation}

\noindent where $Q_{\rm H}$ is thermospheric hydrogen mixing ratio (model top), $H$ is the atmospheric scale height $kt/mg$, and $b$ is the binary Brownian diffusion coefficient given by:

\begin{equation}
    b = 6.5 \times 10^7 T_{\rm thermo}^{0.7}
\end{equation}

\noindent where $T_{\rm thermo}$ is the thermospheric temperature of the atmosphere (taken at 100 km altitude).

We find that IHZ planets with Earth-like atmospheric compositions experiencing water loss should be more resilient to desiccation than previously reported. For example, all simulations around inactive stars have ocean survival timescales well above 10 Gyrs (Figure~\ref{fig:survival}), even for those with H$_2$O$_v$ mixing ratios above $3\e{-3}$ mol mol$^{-1}$ (i.e., classical moist greenhouse). With realistic UV SED however, the oceans are predicted to be lost quickly ($< 1$ Gyr) via the molecular diffusion of H to space. This result stands in contrast to previous estimates using diagnostic H mixing ratios to calculate the escape rates, finding much shorter ocean loss timescales across all host star spectral types (see e.g., Figure 5 in \citealt{KopparapuEt2017ApJ}).  Clearly, a careful assessment of a star’s activity level is critical for determining whether planets around M-dwarfs will lose their oceans to space.

\begin{deluxetable}{cccccc}




\tablecaption{Comparison of simulated results with a suite of GCM studies}

\tablenum{2}

\tablehead{\colhead{Study} & \colhead{$T_{\rm eff}$} & \colhead{T$_s$} & \colhead{H$_2$O} & \colhead{$F_{\rm crit}$} & \colhead{} \\ 
\colhead{} & \colhead{(K)} & \colhead{(K)} & \colhead{(mol mol$^{-1}$)} & \colhead{($F_\oplus)$} & \colhead{} } 

\startdata
K16 & 2600 & N/A & N/A & ${\sim} 1.2$ \\
K16 & 3000 & N/A & N/A & ${\sim}1.4$ \\
K16 & 3300 & 276 & $4 \times 10^{-5}$ & 1.65 \\
K16 & 4000 & 295 & $6 \times 10^{-3}$ & 1.9 \\
K17 & 2600 & 301 & $6 \times 10^{-5}$ & 1.0 \\
K17 & 3000 & 280 & $5 \times 10^{-4}$ & 1.15 \\
K17 & 3300 & 294 & $7 \times 10^{-3}$ & 1.25 \\
K17 & 4000 & 303 & $1 \times 10^{-2}$ & 1.5 \\
Bin18 & 2550 & 285 & $2.1 \times 10^{-3}$ & 0.9 \\
Bin18 & 3050 & 279 & $1.1 \times 10^{-4}$ & 1.0 \\
Bin18 & 3290 & 288 & $9 \times 10^{-4}$ & 1.15 \\
Bin18 & 3960 & 309 & $3.9 \times 10^{-3}$ & 1.35 \\
This Study & 2600 & 282 & $5 \times 10^{-6}$ & 1.0 \\
This Study & 3000 & 278 & $7 \times 10^{-6}$ & 1.1 \\
This Study & 3300 & 298 & $2 \times 10^{-5}$ & 1.6 \\
This Study & 4000 & 301 & $1 \times 10^{-2}$ & 1.9 \\
\enddata


\tablecomments{Approximate values of stellar effective temperature ($T_{\rm eff}$), planetary surface temperature (T$_s$), mixing ratio of stratospheric water vapor (H$_2$O), maximum allowed incident stellar flux before the onset of an incipient runaway greenhouse ($F_{\rm crit}$) across four studies. Bin18 represents CAM5 simulations in \citet{BinEt2018EPSL}, while K16 and K17 represent models of \citet{KopparapuEt2016ApJ} and \citet{KopparapuEt2017ApJ} respectively. H$_2$O mixing ratios are reported at 10 mbar by \citet{BinEt2018EPSL}, 3 mbar by \citet{KopparapuEt2016ApJ}, and 1 mbar by \citet{KopparapuEt2017ApJ} and this study. }


\end{deluxetable}

\hfill \break

\section{Discussion}

This study builds upon previous efforts to study planets near the IHZ, but with the added complexity of interactive 3D photochemistry and atmospheric chemistry, and by self-consistently simulating the atmosphere into the lower thermosphere ($5\e{-6}$ mbar). In comparison with studies that employed self-consistent stellar flux-orbital period relationships, our runaway greenhouse limits are further out (from the respective host stars) than those of \citet{KopparapuEt2016ApJ}, but closer in than those of \citet{KopparapuEt2017ApJ} and \citet{BinEt2018EPSL}. For example, \citet{KopparapuEt2016ApJ} found that the critical flux threshold for thermally stable simulation orbiting a 3000 K star occurs at ($F_{\rm crit}$) ${\sim} 1.3 F_\oplus$, which is approximately 0.2 $F_\oplus$ higher than predicted in this study (Table 2). This discrepancy may be attributable to (i) the inclusion of ozone and its radiative effects in WACCM and (ii) the presence of non-condensable greenhouse gas species. While \citet{KopparapuEt2016ApJ}  include 1 bar of N$_2$ plus 1 ppm of CO$_2$, this study includes additional modern Earth-like CH$_4$ and N$_2$O concentrations, thus yielding IHZ limits that are further away from the host star in comparison to \citet{KopparapuEt2016ApJ}. Note that both studies use the same radiative transfer scheme, cloud physics, and convection scheme. Simulated climates around stars with higher $T_{\rm eff}$ show much smaller differences stemming from lack of ozone heating and reduced degree of inversion, leading to comparable Bond albedos at the inner edge. However, greater disparities are found between our study and \citet{KopparapuEt2017ApJ}. For example, runaway greenhouse occurs at fluxes ($F_{\rm crit}$) ${\sim}0.35 F_\oplus$ higher for simulations across nearly all M-class spectral types (Table 2). This is explained by the finer spectral resolution in the IR and updated H$_2$O absorption by \citet{Wolf+Toon2013Asbio} and \citet{KopparapuEt2017ApJ}, which cause the stratosphere to warm and moisten substantially at a much lower stellar flux. Further, the native radiative transfer of CAM4 is shown to be too weak, both in the longwave and shortwave, with respect to water vapor absorption \citep{YangEt2016ApJ}. Differences between previous GCM calculations of the IHZ around Sun-like stars (e.g., CAM4; \citealt{Wolf+Toon2015JGR} and LMD: \citealt{LeconteEt2013NATURE}) can also be attributed to treatment of moist physics and clouds \citep{YangEt2019ApJb}.

Our predictions of water loss and habitability implications show greater divergence from previous GCM studies$-$an outcome that is not unexpected given different initial atmospheric compositions and the addition of model chemistry. For quiescent stars, model top H mixing ratio predictions (hence water loss rates) presented here are orders of magnitude lower than previous work with simplified atmospheric compositions and without interactive chemistry. Implications of our results are favorable to the survival of surface liquid water for planets around quiescent M-dwarfs. For example, a recent study of the temporal radiation environment of the LHS 1140 system suggests that the planet receives relatively constant NUV ($177 - 283$ nm) flux $<2\%$ compared to that of the Earth \citep{SpenelliEt2019arXiv}. Our results suggest that LHS 1140b is likely stable against complete ocean desiccation due to the low UV activity of the host star, which bodes well for its habitability. Note however, that since our WACCM simulations assume a hydrostatic atmosphere, escape of H$_2$O$_v$ is only roughly approximated. Furthermore, during the super-luminous pre-main sequence stages of M-dwarfs (<100 Myr), high amounts of X-ray/EUV irradiation may cause an early desiccation and/or runway greenhouse of planetary atmospheres in the IHZ \citep{Luger+Barnes2015AsBio}. Even so, rocky planets around M-dwarfs may still possess active hydrological cycles through acquisition of cometary materials \citep{Tian+Ida2015NatGeo} as well as extended deep mantle cycling and the emergence of secondary atmospheres \citep{Komacek+Abbot2016ApJ}. Despite the super-luminous stages of M-dwarfs, existence of abundant water inventories is shown to be plausible using numerical TTV analysis, for example, in the TRAPPIST-1 system \citep{GrimmEt2018A&A}. For this pilot CCM study of the IHZ, we focus on main-sequence stars to be consistent with previous work modeling moist greenhouse states (e.g., \citealt{KastingEt1993Icarus,KopparapuEt2017ApJ}). Further study is warranted examining stellar activity levels, including enhanced UV flux, time-dependent stellar flares, and sun-like proton events, and their roles in driving water-loss in habitable planet atmospheres.

Coupled CCMs, such as the one employed here, are advantageous for helping to improve/inform 1D model simulations. Previous work (e.g., \citealt{Zhang&Showman2018ApJ}) have shown that the constant vertical diffusion coefficients assumed in 1D models (e.g., \citealt{HuEt2012ApJ,Kaltenegger+2010ApJ}) may be invalid for different chemical compounds. Here, we find that the efficiency of global-mean vertical transport is not only species-dependent, but also host star dependent, as the magnitude of meridional overturning circulation and degree of vertical wave mixing are inherently tied to the planetary rotation rate and stellar $T_{\rm eff}$ (Figure~\ref{fig:oz_trans}), both of which are constrained for synchronously rotating planets.  Although a detailed comparison of the full set of our chemical constituent profiles with those of 1D is beyond the scope of this study, our results show that 3D CCMs could offer a basis to improve prediction of the 1D vertical distribution of photochemically important species (e.g., ozone). This task is especially important in the transition regime (i.e., for planets around stars with $2900 \la T_{\rm eff} \la 3400$ K), where stratospheric circulation patterns can shift substantially (i.e., emergence of anti-Brewer-Dobson cells; \citealt{CaroneEt2018MNRAS}), leading to the further breakdown of a fixed vertical diffusivity assumption by 1D models.

In this study we focus solely on simulations with Earth-like ocean coverage and landmass distributions. However, water inventories vary with accretion and escape history. If a planet is barren (i.e., without a substantial surface liquid water inventory), then moist convection is inhibited and the water vapor greenhouse effect is suppressed. This can result in the delay of a runaway greenhouse \citep{AbeEt2011AsBio} and the onset of moist bistability wherein surface water could condense in colder reservoirs \citep{LeconteEt2013A&A}. A similar effect could occur if the substellar point is located above a large landmass instead of an ocean basin \citep{LewisEt2018ApJ}. This has relevance for atmospheric chemistry and habitability as it could suppress substellar moisture, leading to lower production rates of H in the thermosphere  (above $10^{-2}$ mbar) and OH in the stratosphere (between 100 and 1 mbar). 

Similarly, we fix our substellar point over an ocean basin and assume circular orbits locked in 1:1 spin-orbit resonance. In reality, the substellar point and stellar zenith angle could be nonstationary \citep{LeconteEt2015SCI} and planetary orbits could be eccentric without the stabilizing influence of a gas giant \citep{TsiganisEt2005NAT}. Nonstationary solar zenith angles could affect atmospheric circulation by modulating efficiency of moist convection, while eccentricity could drive planetary climate by as evidenced by Earth’s geological record \citep{HortonEt2012PPP}. However, these considerations are arguably secondary as existence of thermal tides are theoretical in the context of exoplanets and planets in the RV samples that have eccentricity greater than 0.1 are not common \citep{ShenEt2008ApJ}. Thus, we believe our simplification of fixed substellar point and perfect circular orbit should be valid for the majority of actual planetary systems.

Apart from surface climate, continued habitability is contingent upon the formation and retention of an ozone layer to shield excessive stellar UV-C ($200 < \lambda < 280$ nm) radiation and energetic particle bombardment. A thin ozone layer is hazardous to DNA due to surface exposure to high doses of UV radiation (e.g, \citealt{OMalley+Kaltenegger2017MNRAS}). Alternatively, UV radiation may be critical in instigating complex prebiotic chemistry (e.g., \citealt{RanjanEt2017ApJ}). As our simulations enter the moist greenhouse regime, we find that their atmospheres have orders of magnitude lower ozone mixing ratios than those in temperate climates (Figure~\ref{fig:flux}g),  implying that the UV fluxes reaching the planetary surface may be high and therefore potentially threatening to surface life.   Further, we find that both stellar UV activity and efficiency of day-to-nightside ozone transport could control the degree of UV flux penetration on the dayside surface. Thus, future constraints on the width of this “complex life habitable zone” (HZCL; \citealt{SchwietermanEt2019ApJ}) will need to evaluate its dependencies on stellar flux, spectral type, and stellar activity, and will benefit from the use 3D CCMs.

Ultimately, CCM predictions of planetary habitability near the IHZ depend on the water accretion history \citep{RaymondEt2006Icarus}, stellar XUV evolution  \citep{Luger+Barnes2015AsBio}, orbital parameters \citep{KilicEt2017ApJ}, and the spatial distribution of surface water \citep{WayEt2016GRL,KodamaEt2018JGR}. These considerations should be investigated in a more extensive CCM-based parameter space study (similar in spirit to e.g., \citealt{Komacek+Abbot2019ApJ}) to better understand the climate, chemistries, and habitability potentials of IHZ planets around M-dwarfs.

\begin{figure*}[t] 
\begin{center}
\includegraphics[width=1.65\columnwidth]{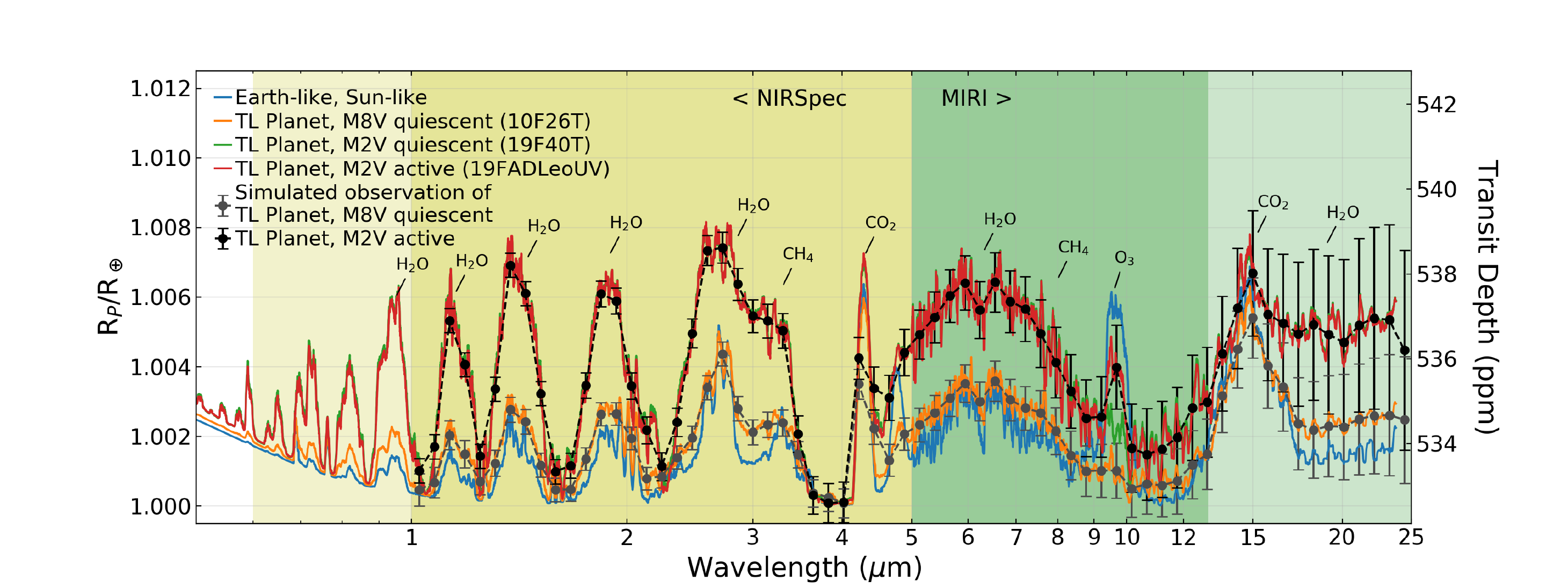}
\includegraphics[width=1.65\columnwidth]{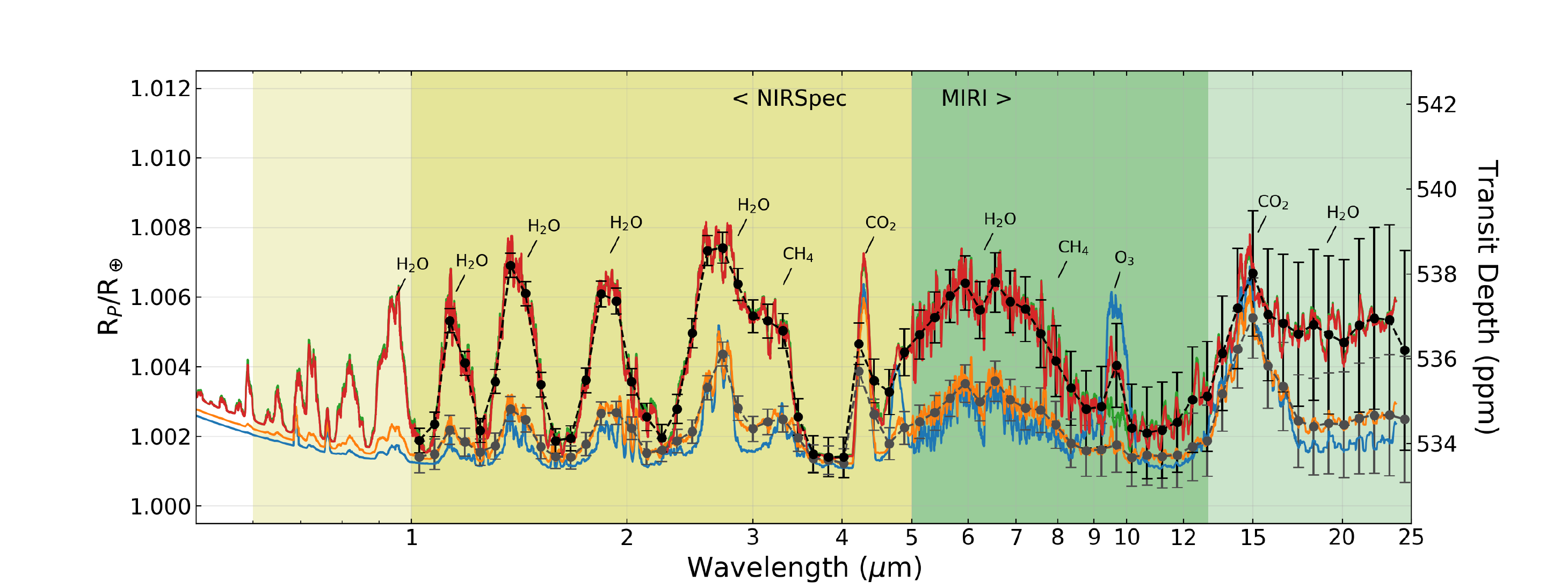}
\includegraphics[width=1.65\columnwidth]{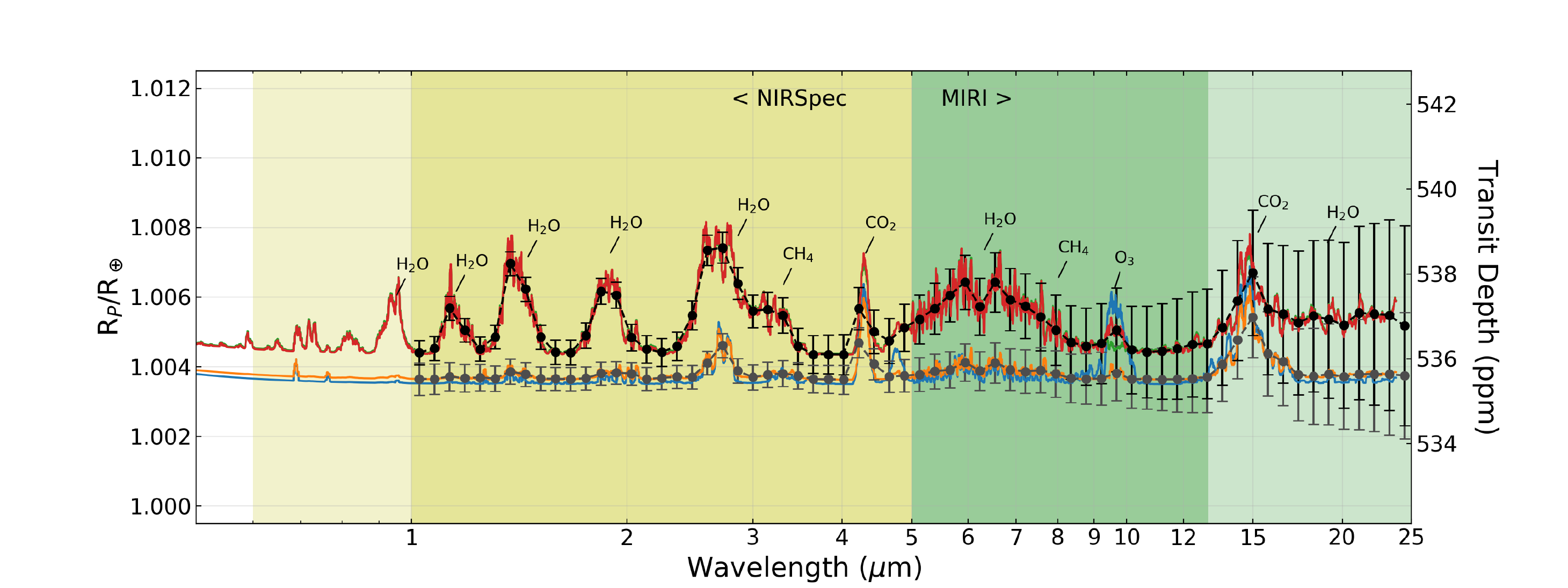}
\includegraphics[width=1.65\columnwidth]{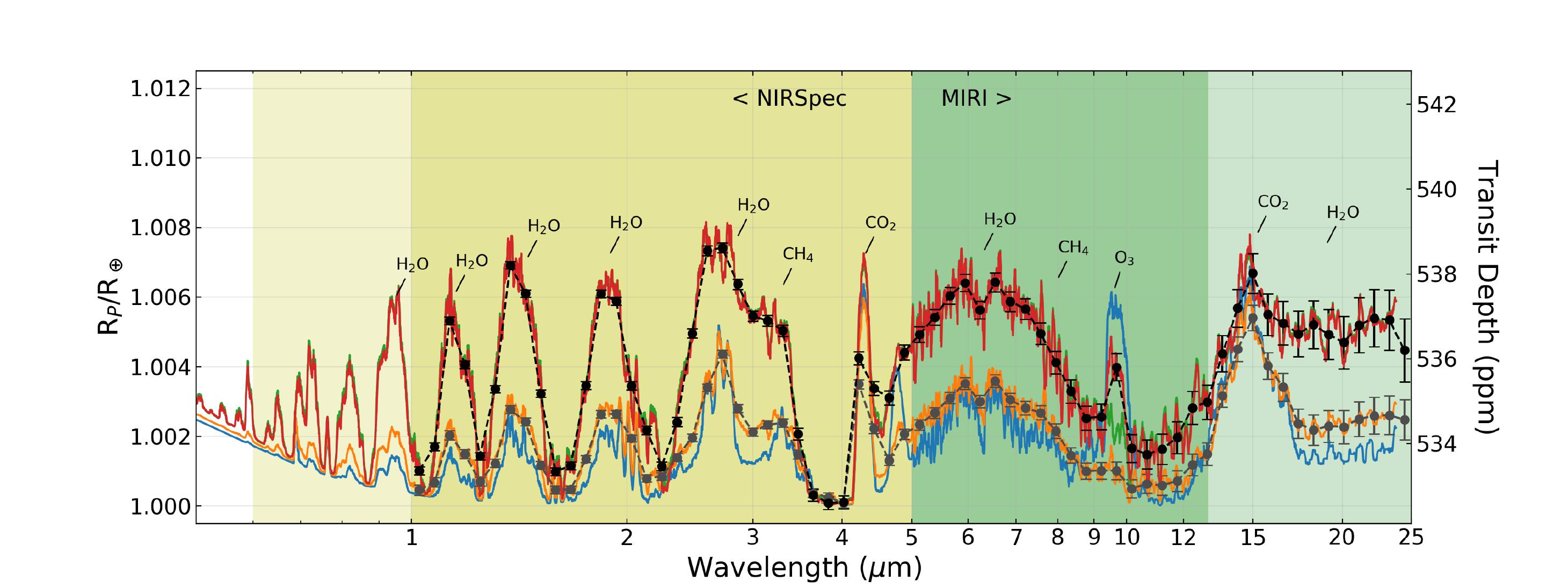}
\caption{\label{fig:obs1}  Simulated atmosphere transmission spectra of synchronously-rotating planets around M-stars and an Earth-like fast rotator around a Sun-like star (P = 24 hrs), showing apparent planet radius and transit depth as a function of wavelength ($\mu$m). Simulated input data from CCM (i.e., experiments 10F26T, 19F40T, and 19FADLeoUV) are spatially averaged across the terminators. We explore three cloud assumptions: cloudless (top and bottom panels), uniform grey cloud at 100 mbar with 0.5 opacity (second panel), and uniform grey cloud at 10 mbar with 1.0 opacity (third panel). Also shown are simulated JWST observation with $1\sigma$ uncertainty bar (black) at two integration times: 10 hr (first three panels) and 100 hr (bottom panel). The simulated observation assumes the planet to be 2 pc away from the observer, and the bin width of the telescope to be 1 $\mu$m at wavelength of 10 $\mu$m.  } 
\end{center}
\end{figure*}

\subsection{Observational Implications \& Detectability}
Follow-up characterization efforts by future instruments will likely target planets around M-dwarfs. To contextualize our CCM results within an observational framework, we calculate transmission spectra, secondary eclipse thermal emission spectra, and their simulated observations using the Simulated Exoplanet Atmosphere Spectra (SEAS) model (Zhan et al. in revision). SEAS is a radiative transfer code that calculates the attenuation of photons by molecular absorption and Rayleigh/Mie scattering as the photons travel through a hypothetical exoplanet atmosphere. The simulation approach is similar to previous work by \citet{KemptonEt2017ApJL} and \citet{KemptonEt2009ApJ}. The molecular absorption cross-section for O$_2$, H$_2$O, CO$_2$, CH$_4$, O$_3$, and H are calculated using the HITRAN2016 molecular line-list database \citep{GordonEt2017}. The SEAS transmission spectra are validated through comparison of its simulated Earth transmission spectrum with that of real Earth counterparts measured by the Atmospheric Chemistry Experiment (ACE) data set \citep{BernathEt2015GRL}. For more details on SEAS, please see Section 3.4 of Zhan et al., (in revision). 

To compute atmospheric spectra, we use a subset of our CCM results: (i) Earth around the Sun, (ii) tidally-locked planet around an M8V star (10F26T), (iii) tidally-locked planet around a quiescent M2V star (19F40T), and (iv) tidally-locked planet around an active M2V star (19FADLeoUV). These simulations are chosen to illustrate the spectral feature differences between rapidly-rotating planets (10F26T and Earth-Sun) and slowly-rotating planets (19F40T and 19FADLeoUV). These simulations also demonstrate the consequences of different stellar UV activity levels on the spectral shapes: from low (10F26T and 19F40T) to mid (Earth-Sun), to high (19FADLeoUV) UV inputs. Lastly, planets orbiting late K-dwarfs, such as 19F40T, are argued to be at an advantage over those around mid-to-late M-dwarfs for biosignature potential \citep{Arney2019ApJL,Lingam+Loeb2017ApJL,Lingam+Loeb2018JCAP}, and thus our primary focus on the host stars with $T_{\rm eff}$ of 4000 K.

CCM inputs for the SEAS model include: simulated temperatures and mixing ratios of gaseous constituents (i.e., N$_2$, CO$_2$, H$_2$O$_v$, O$_2$, O$_3$, CH$_4$, and N$_2$O), converted to 1D vertical time-averaged profiles. Transmission spectra were generated using the terminator mean values, while the emission spectra used the dayside-mean. SEAS assumes the premise of clouds, rather than using CCM results, in order to facilitate comparison with previous work \citep{MorleyEt2013ApJ,HengEt2016ApJL}. For transmission, we explore three scenarios: uniform grey cloud at 10 mbar with 1.0 opacity (or optical depth), uniform grey cloud at 100 mbar with 0.5 opacity, and no clouds. For thermal emission, we assumed a 50\% patchy grey cloud at 10 mbar with 0.5 opacity. These selections are made based on the fact that the atmosphere molecular absorption path length for stellar radiation passing through the rim of the planet atmosphere is ${\sim} 10\times$ the molecular absorption pathlength for blackbody radiation traveling from the surface of the planet. Parameterized clouds are used rather than CCM simulated clouds, as the former can set an upper/lower bound to our detection threshold and thus facilitate comparison with previous work. For instance, the 10 mbar with 1.0 opacity case mimics that of an ``upper atmosphere uniform haze", e.g., \citet{Kawashima+Ikoma2018ApJ}. Moreover, GCM simulation of clouds remains an active area of research and thus simulated clouds come with inherent uncertainties. In future efforts, we will endeavor to include simulated clouds and their inherent uncertainties from a suite of GCMs into CCM-SEAS. 

We validated our simulated Earth atmosphere transmission spectra with measurements of Earth’s atmosphere through the ACE program \citep{BernathEt2015GRL} and emission spectra with MODTRAN \citep{Berk2014SPIE}. Minor differences are due to exclusion of trace gases in the Earth atmosphere with a column average mixing ratio less than 1 ppmv.

Our simulated observations assume that the system is 2 pc from the observer. The planet of consideration is an Earth-sized and Earth mass planet. We also assume the use of a JWST-like, 6.5 m telescope, and with 25\% throughput and an approximated noise multiplier of 50\% which accounts for potentially unknown stellar variability and/or instrumental effects. While the spectral resolution of JWST is R = 100 at 1 - 5 $\mu$m (NIRSpec) and R = 160 at 5 - 12 $\mu$m (MIRI), in practice this ``high" resolution (as compared to Hubble WFC3) is not prioritized for detection/distinction of H$_2$O$_v$ and O$_3$ molecules in exoplanet atmospheres due to the unique and broad spectra features these two molecules have (Zhan, et al. submitted). Therefore, detection can be optimized by using a larger bin width to increase the signal-to-noise ratio (SNR) of the H$_2$O$_v$ and O$_3$ molecules at the expense of reducing the resolution just enough to distinguish the molecules in consideration. We use the empirical formula of:

\begin{equation}
l_n = l_0\left(\frac{\lambda_n}{\lambda_0}\right)^m
\end{equation}

\noindent where $l_0$ is 0.1 $\mu$m and $\lambda_0$ is 1 $\mu$m. i.e, the bin width $l_n$, at $\lambda_n$ = 10 $\mu$m, is = 1 $\mu$m. Note that we neglect the systematic noise of JWST, which is projected to be on the order of ${\sim}10$ ppm \citep{GreeneEt2016ApJ}. Implementation of the noise floor into the JWST simulator will slightly weaken our predicted features. Given the planet is at 2 pc and the high stratospheric H$_2$O concentration however, the H$_2$O features will likely still be detectable.

We find that the detection of H$_2$O$_v$ in moist greenhouse atmospheres can be achieved with high SNR confidence using NIRSpec. We also find that O$_3$ detection at the 9.6 $\mu$m window can potentially reach an SNR of 3 using MIRI LRS for planets around active stars (e.g., a Sun-like star or active M-dwarf). For the transmission spectra (Figure~\ref{fig:obs1}\footnote{The planet radius is independent of planetary system, and is easier for comparison with other theory work, as ppm depend on planet/star radius ratio.}), we compare our simulated results with those of \citet{FujiiEt2017ApJ} and \citet{KopparapuEt2017ApJ}, and find that our CCM-SEAS results are in agreement with the potential of characterizing water vapor features between 2.5 and 8 $\mu$m.

Slowly-rotating planets orbiting the IHZ of M2V stars (e.g., 19F40T and 19FADLeoUV) have moist greenhouse atmospheres and higher stratospheric water vapor content (red curves in Figure 5), which raises their signals in comparison to rapidly-rotating planets and stratospherically dry planets (e.g., 10F26T). In a similar vein, detection of water vapor can be achieved at high SNR confidence only for the moist greenhouse case, where H$_2$O$_v$ is four orders of magnitude more abundant than in the Earth's stratosphere. Detection of water vapor is difficult for an Earth-like atmosphere in transmission as the majority of water vapor is concentrated below the tropopause. Transmission spectra of IHZ planets generated with 1D models (e.g., \citet{LincowskiEt2018ApJ}) do not display these prominent water vapor features as large-scale hydrological circulation processes are not resolved in 1D. We also find that the results of parameterized clouds match those of previous work \citep{StevensonEt2016ApJL,WakefordEt2019arXiv} such that they inhibit detection of molecular features (Figure~\ref{fig:obs1}).

Oxygenated-atmospheres around active stars (i.e., Earth and 19FADLeoUV) should have more pronounced ozone features due to increased ozone production rates compared to their quiescent counterparts. We find that although the H$_2$O$_v$ features are not significantly altered by stellar UV activity, the O$_3$ features are. In addition, we test the effects of total integration time (10 hr vs 100 hr) on the predicted observations to explore the ``most optimistic" scenario (Figure 11). These integration times translate to 4 to 7 and 40 to 70 transits respectively for a typical M-dwarf system. We find that detectability of the ozone feature is substantially improved with a higher integration time (Figure~\ref{fig:obs1}).

For secondary eclipse thermal emission spectra, we find that the 9.6 $\mu$m O$_3$ feature is located near the emission peak of the planet (${\sim}300$ K blackbody). Despite this finding, detection of this feature via secondary eclipse could be challenging for Earth-sized planets near 4000 K stars due to low SNR confidence.  Potentially low SNRs are a result of the constraints of JWST’s cryogenic lifetime (necessary for mid-IR observations) of 5 years. Further, as the total number of transit hours that can accumulate is less than 100, the maximum achievable SNR confidence given our simulation parameters would be less than 3. For a super-Earth (e.g. $1.75~R_\oplus$) or a hotter (but non-habitable) planet around a late M-dwarf star however, secondary eclipse measurements of these features could be achievable \citep{KollEt2019arXiv,MalikEt2019arXiv}.

\section{Conclusion}
\label{sec:conclusion}

In this study, we carried out numerical simulations of climate and chemistries of tidally-locked planets with a 3D CCM. Our results show that the maintenance of ozone layers, water photodissociation efficiency, and the onset of moist and incipient runaway greenhouse states depend on the incident stellar flux, stellar spectral-type, and importantly, UV radiation. By directly simulating photochemically important species such as ozone, we find that their abundance and distribution depend on the host star spectral type.  The strength of the stratospheric overturning circulation, for example, increases with stellar $T_{\rm eff}$, leading to higher efficiency in the divergent transport of airmasses and thus photochemically produced species and aerosols.

Critically, we find that only climates around active M-dwarfs enter the classical moist greenhouse regime, wherein hydrogen mixing ratios are sufficiently high such that water loss could evaporate the surface ocean within 5 Gyrs. For those around quiescent M-dwarfs, hydrogen mixing ratios do not exceed that of water vapor. As a consequence, we find that planets orbiting quiescent stars have much longer ocean survival timescales than those around active M-dwarfs. Thus, our results suggest that improved constraints on the UV activity of low-mass stars will be critical in understanding the long-term habitability of future discovered exoplanets (e.g., in the TESS sample; \citealt{GuntherEt2019arXiv}).

Stellar UV radiation has pronounced effects on atmospheric circulation and chemistry. Our 3D CCM simulations show that vertical and horizontal winds in the upper atmosphere (${\sim}1$ mbar) are strengthened with higher UV fluxes. Global distributions of O$_3$, OH, and H are the result of long-term averaged tradeoffs between dynamical, photolytic, and photochemical processes$-$resulting in substantially different day-to-nightside contrasts with incident UV radiation. Thus, coupling dynamics and photochemistry will be necessary to better understand the spatial distributions and temporal variability (e.g., \citealt{OlsonEt2018ApJL}) of biogenic compounds and their byproducts.

Using a radiative transfer model with our CCM results as inputs, we show that detecting prominent water vapor and ozone features on M-dwarf planets during primary transits is possible by future instruments such as the JWST \citep{BeichmanEt2014PASP}. However, secondary eclipse observations are more challenging due to the predicted low SNR confidence.

\acknowledgements
H.C. thanks the R. K. Kopparapu, S. D. Domagal-Goldman, and the Exoplanet Journal Club at the University of Chicago for stimulating discussions as well as J. Yang and M. Lingam for helpful comments. H.C. and D.E.H. acknowledge support from the Future Investigators in NASA Earth and Space Science and Technology (FINESST) Graduate Research Award 80NSSC19K1523. E.T.W. thanks NASA Habitable Worlds Grant 80NSSC17K0257 for support. Z.Z. thanks the MIT BOSE Fellow program, the Change Happens Foundation, and the Heising-Simons Foundation for partial funding of this work. We acknowledge and thank the computational, storage, data analysis, and staff resources provided by the QUEST high performance computing facility at Northwestern University, which is jointly supported by the Office of the Provost, Office for Research, and Northwestern University Information Technology.

\bibliographystyle{apj}

\begin{thebibliography}{}
\expandafter\ifx\csname natexlab\endcsname\relax\def\natexlab#1{#1}\fi

\bibitem[{{Abbot} {et~al.}(2012){Abbot}, {Cowan}, \& {Ciesla}}]{AbbotEt2012ApJ}
{Abbot}, D.~S., {Cowan}, N.~B., \& {Ciesla}, F.~J. 2012, \apj, 756, 178

\bibitem[{{Abe} {et~al.}(2011){Abe}, {Abe-Ouchi}, {Sleep}, \&
  {Zahnle}}]{AbeEt2011AsBio}
{Abe}, Y., {Abe-Ouchi}, A., {Sleep}, N.~H., \& {Zahnle}, K.~J. 2011,
  Astrobiology, 11, 443

\bibitem[{{Anglada-Escud{\'e}} {et~al.}(2016){Anglada-Escud{\'e}}, {Amado},
  {Barnes}, {Berdi{\~n}as}, {Butler}, {Coleman}, {de La Cueva}, {Dreizler},
  {Endl}, {Giesers}, {Jeffers}, {Jenkins}, {Jones}, {Kiraga}, {K{\"u}rster},
  {L{\'o}pez-Gonz{\'a}lez}, {Marvin}, {Morales}, {Morin}, {Nelson}, {Ortiz},
  {Ofir}, {Paardekooper}, {Reiners}, {Rodr{\'{\i}}guez},
  {Rodr{\'{\i}}guez-L{\'o}pez}, {Sarmiento}, {Strachan}, {Tsapras}, {Tuomi}, \&
  {Zechmeister}}]{AngladaEt2016NATURE}
{Anglada-Escud{\'e}}, G., {Amado}, P.~J., {Barnes}, J., {et~al.} 2016, Nature,
  536, 437

\bibitem[{{Arney}(2019)}]{Arney2019ApJL}
{Arney}, G.~N. 2019, \apjl, 873, L7

\bibitem[{Banks \& Kockarts(1973)}]{BanksET1973}
Banks, P., \& Kockarts, G. 1973, Aeronomy, vol. B, chap. 15

\bibitem[{{Barclay} {et~al.}(2018){Barclay}, {Pepper}, \&
  {Quintana}}]{BarclayEt2018ApJS}
{Barclay}, T., {Pepper}, J., \& {Quintana}, E.~V. 2018, The Astrophysical
  Journal Supplement Series, 239, 2

\bibitem[{Bardeen {et~al.}(2010)Bardeen, Toon, Jensen, Hervig, Randall, Benze,
  Marsh, \& Merkel}]{BardeenEt2010JGR}
Bardeen, C., Toon, O., Jensen, E., {et~al.} 2010, Journal of Geophysical
  Research: Atmospheres, 115

\bibitem[{{Beichman} {et~al.}(2014){Beichman}, {Benneke}, {Knutson}, {Smith},
  {Lagage}, {Dressing}, {Latham}, {Lunine}, {Birkmann}, {Ferruit}, {Giardino},
  {Kempton}, {Carey}, {Krick}, {Deroo}, {Mandell}, {Ressler}, {Shporer},
  {Swain}, {Vasisht}, {Ricker}, {Bouwman}, {Crossfield}, {Greene}, {Howell},
  {Christiansen}, {Ciardi}, {Clampin}, {Greenhouse}, {Sozzetti}, {Goudfrooij},
  {Hines}, {Keyes}, {Lee}, {McCullough}, {Robberto}, {Stansberry}, {Valenti},
  {Rieke}, {Rieke}, {Fortney}, {Bean}, {Kreidberg}, {Ehrenreich}, {Deming},
  {Albert}, {Doyon}, \& {Sing}}]{BeichmanEt2014PASP}
{Beichman}, C., {Benneke}, B., {Knutson}, H., {et~al.} 2014, \pasp, 126, 1134

\bibitem[{{Benneke} {et~al.}(2019{\natexlab{a}}){Benneke}, {Knutson},
  {Lothringer}, {Crossfield}, {Moses}, {Morley}, {Kreidberg}, {Fulton},
  {Dragomir}, {Howard}, {Wong}, {D{\'e}sert}, {McCullough}, {Kempton},
  {Fortney}, {Gilliland}, {Deming}, \& {Kammer}}]{BennekeEt2019NAtAstro}
{Benneke}, B., {Knutson}, H.~A., {Lothringer}, J., {et~al.} 2019{\natexlab{a}},
  Nature Astronomy, arXiv:1907.00449

\bibitem[{{Benneke} {et~al.}(2019{\natexlab{b}}){Benneke}, {Wong}, {Piaulet},
  {Knutson}, {Crossfield}, {Lothringer}, {Morley}, {Gao}, {Greene}, {Dressing},
  {Dragomir}, {Howard}, {McCullough}, {Fortney}, \&
  {Fraine}}]{BennekeEt2019arXiv}
{Benneke}, B., {Wong}, I., {Piaulet}, C., {et~al.} 2019{\natexlab{b}}, arXiv
  e-prints, arXiv:1909.04642

\bibitem[{{Berk} {et~al.}(2014){Berk}, {Conforti}, {Kennett}, {Perkins},
  {Hawes}, \& {van den Bosch}}]{Berk2014SPIE}
{Berk}, A., {Conforti}, P., {Kennett}, R., {et~al.} 2014, in \procspie, Vol.
  9088, Algorithms and Technologies for Multispectral, Hyperspectral, and
  Ultraspectral Imagery XX, 90880H

\bibitem[{Bernath {et~al.}(2005)Bernath, McElroy, Abrams, Boone, Butler,
  Camy-Peyret, Carleer, Clerbaux, Coheur, Colin, {et~al.}}]{BernathEt2015GRL}
Bernath, P.~F., McElroy, C.~T., Abrams, M., {et~al.} 2005, Geophysical Research
  Letters, 32

\bibitem[{{Bin} {et~al.}(2018){Bin}, {Tian}, \& {Liu}}]{BinEt2018EPSL}
{Bin}, J., {Tian}, F., \& {Liu}, L. 2018, Earth and Planetary Science Letters,
  492, 121

\bibitem[{{Carone} {et~al.}(2015){Carone}, {Keppens}, \&
  {Decin}}]{CaroneEt2015MNRAS}
{Carone}, L., {Keppens}, R., \& {Decin}, L. 2015, \mnras, 453, 2412

\bibitem[{{Carone} {et~al.}(2018){Carone}, {Keppens}, {Decin}, \&
  {Henning}}]{CaroneEt2018MNRAS}
{Carone}, L., {Keppens}, R., {Decin}, L., \& {Henning}, T. 2018, \mnras, 473,
  4672

\bibitem[{{Chen} {et~al.}(2018){Chen}, {Wolf}, {Kopparapu}, {Domagal-Goldman},
  \& {Horton}}]{ChenEt2018ApJL}
{Chen}, H., {Wolf}, E.~T., {Kopparapu}, R., {Domagal-Goldman}, S., \& {Horton},
  D.~E. 2018, \apjl, 868, L6

\bibitem[{Collins {et~al.}(2006)Collins, Bitz, Blackmon, Bonan, Bretherton,
  Carton, Chang, Doney, Hack, Henderson, {et~al.}}]{CollinsEt2006}
Collins, W.~D., Bitz, C.~M., Blackmon, M.~L., {et~al.} 2006, Journal of
  Climate, 19, 2122

\bibitem[{{Del Genio} {et~al.}(2019){Del Genio}, {Way}, {Amundsen}, {Aleinov},
  {Kelley}, {Kiang}, \& {Clune}}]{DelGenioEt2019AsBio}
{Del Genio}, A.~D., {Way}, M.~J., {Amundsen}, D.~S., {et~al.} 2019,
  Astrobiology, 19, 99

\bibitem[{{del Genio} {et~al.}(1993){del Genio}, {Zhou}, \&
  {Eichler}}]{DelGenioEt1993Icar}
{del Genio}, A.~D., {Zhou}, W., \& {Eichler}, T.~P. 1993, \icarus, 101, 1

\bibitem[{{Dittmann} {et~al.}(2017){Dittmann}, {Irwin}, {Charbonneau},
  {Bonfils}, {Astudillo-Defru}, {Haywood}, {Berta-Thompson}, {Newton},
  {Rodriguez}, {Winters}, {Tan}, {Almenara}, {Bouchy}, {Delfosse}, {Forveille},
  {Lovis}, {Murgas}, {Pepe}, {Santos}, {Udry}, {W{\"u}nsche}, {Esquerdo},
  {Latham}, \& {Dressing}}]{DittMannEt2017NATURE}
{Dittmann}, J.~A., {Irwin}, J.~M., {Charbonneau}, D., {et~al.} 2017, Nature,
  544, 333

\bibitem[{{Dragomir} {et~al.}(2019){Dragomir}, {Teske}, {G{\"u}nther},
  {S{\'e}gransan}, {Burt}, {Huang}, {Vanderburg}, {Matthews}, {Dumusque},
  {Stassun}, {Pepper}, {Ricker}, {Vanderspek}, {Latham}, {Seager}, {Winn},
  {Jenkins}, {Beatty}, {Bouchy}, {Brown}, {Butler}, {Ciardi}, {Crane},
  {Eastman}, {Fossati}, {Francis}, {Fulton}, {Gaudi}, {Goeke}, {James},
  {Klaus}, {Kuhn}, {Lovis}, {Lund}, {McDermott}, {Paegert}, {Pepe},
  {Rodriguez}, {Sha}, {Shectman}, {Shporer}, {Siverd}, {Garcia Soto},
  {Stevens}, {Twicken}, {Udry}, {Villanueva}, {Wang}, {Wohler}, {Yao}, \&
  {Zhan}}]{DragomirEt2019ApJL}
{Dragomir}, D., {Teske}, J., {G{\"u}nther}, M.~N., {et~al.} 2019, \apjl, 875,
  L7

\bibitem[{{Dressing} \& {Charbonneau}(2015)}]{Dressing+Charbonneau2015ApJ}
{Dressing}, C.~D., \& {Charbonneau}, D. 2015, The Astrophysical Journal
  Letters, 807, 45

\bibitem[{{Edson} {et~al.}(2011){Edson}, {Lee}, {Bannon}, {Kasting}, \&
  {Pollard}}]{EdsonEt2011Icar}
{Edson}, A., {Lee}, S., {Bannon}, P., {Kasting}, J.~F., \& {Pollard}, D. 2011,
  \icarus, 212, 1

\bibitem[{Fomichev {et~al.}(1998)Fomichev, Blanchet, \&
  Turner}]{FomichevE1998JGR}
Fomichev, V., Blanchet, J.-P., \& Turner, D. 1998, Journal of Geophysical
  Research: Atmospheres, 103, 11505

\bibitem[{{France} {et~al.}(2013){France}, {Froning}, {Linsky}, {Roberge},
  {Stocke}, {Tian}, {Bushinsky}, {D{\'e}sert}, {Mauas}, {Vieytes}, \&
  {Walkowicz}}]{FranceEt2013ApJ}
{France}, K., {Froning}, C.~S., {Linsky}, J.~L., {et~al.} 2013, \apj, 763, 149

\bibitem[{{Fujii} {et~al.}(2017){Fujii}, {Del Genio}, \&
  {Amundsen}}]{FujiiEt2017ApJ}
{Fujii}, Y., {Del Genio}, A.~D., \& {Amundsen}, D.~S. 2017, \apj, 848, 100

\bibitem[{Gill(1980)}]{Gill1980}
Gill, A.~E. 1980, Quarterly Journal of the Royal Meteorological Society, 106,
  447

\bibitem[{{Gillon} {et~al.}(2017){Gillon}, {Triaud}, {Demory}, {Jehin}, {Agol},
  {Deck}, {Lederer}, {de Wit}, {Burdanov}, {Ingalls}, {Bolmont}, {Leconte},
  {Raymond}, {Selsis}, {Turbet}, {Barkaoui}, {Burgasser}, {Burleigh}, {Carey},
  {Chaushev}, {Copperwheat}, {Delrez}, {Fernandes}, {Holdsworth}, {Kotze}, {Van
  Grootel}, {Almleaky}, {Benkhaldoun}, {Magain}, \&
  {Queloz}}]{GillonEt2017NATURE}
{Gillon}, M., {Triaud}, A.~H.~M.~J., {Demory}, B.-O., {et~al.} 2017, Nature,
  542, 456

\bibitem[{Gordon {et~al.}(2017)Gordon, Rothman, Hill, Kochanov, Tan, Bernath,
  Birk, Boudon, Campargue, Chance, {et~al.}}]{GordonEt2017}
Gordon, I.~E., Rothman, L.~S., Hill, C., {et~al.} 2017, Journal of Quantitative
  Spectroscopy and Radiative Transfer, 203, 3

\bibitem[{{Greene} {et~al.}(2016){Greene}, {Line}, {Montero}, {Fortney},
  {Lustig-Yaeger}, \& {Luther}}]{GreeneEt2016ApJ}
{Greene}, T.~P., {Line}, M.~R., {Montero}, C., {et~al.} 2016, \apj, 817, 17

\bibitem[{{Grimm} {et~al.}(2018){Grimm}, {Demory}, {Gillon}, {Dorn}, {Agol},
  {Burdanov}, {Delrez}, {Sestovic}, {Triaud}, {Turbet}, {Bolmont}, {Caldas},
  {de Wit}, {Jehin}, {Leconte}, {Raymond}, {Van Grootel}, {Burgasser}, {Carey},
  {Fabrycky}, {Heng}, {Hernandez}, {Ingalls}, {Lederer}, {Selsis}, \&
  {Queloz}}]{GrimmEt2018A&A}
{Grimm}, S.~L., {Demory}, B.-O., {Gillon}, M., {et~al.} 2018, \aap, 613, A68

\bibitem[{{G{\"u}nther} {et~al.}(2019){G{\"u}nther}, {Zhan}, {Seager},
  {Rimmer}, {Ranjan}, {Stassun}, {Oelkers}, {Daylan}, {Newton}, {Gillen},
  {Rappaport}, {Ricker}, {Latham}, {Winn}, {Jenkins}, {Glidden}, {Fausnaugh},
  {Levine}, {Dittmann}, {Quinn}, {Krishnamurthy}, \&
  {Ting}}]{GuntherEt2019arXiv}
{G{\"u}nther}, M.~N., {Zhan}, Z., {Seager}, S., {et~al.} 2019, arXiv e-prints,
  arXiv:1901.00443

\bibitem[{{Hack}(1994)}]{Hack1994JGR}
{Hack}, J.~J. 1994, \jgr, 99, 5551

\bibitem[{{Haqq-Misra} {et~al.}(2018){Haqq-Misra}, {Wolf}, {Joshi}, {Zhang}, \&
  {Kopparapu}}]{Haqq-MisraEt2018ApJ}
{Haqq-Misra}, J., {Wolf}, E.~T., {Joshi}, M., {Zhang}, X., \& {Kopparapu},
  R.~K. 2018, \apj, 852, 67

\bibitem[{{Hart}(1979)}]{Hart1979Icarus}
{Hart}, M.~H. 1979, \icarus, 37, 351

\bibitem[{{Haynes} {et~al.}(1991){Haynes}, {McIntyre}, {Shepherd}, {Marks}, \&
  {Shine}}]{HaynesEt1991}
{Haynes}, P.~H., {McIntyre}, M.~E., {Shepherd}, T.~G., {Marks}, C.~J., \&
  {Shine}, K.~P. 1991, Journal of Atmospheric Sciences, 48, 651

\bibitem[{{Held} \& {Suarez}(1994)}]{Held+Suarez1994}
{Held}, I.~M., \& {Suarez}, M.~J. 1994, Bulletin of the American Meteorological
  Society, 75, 1825

\bibitem[{{Heng}(2016)}]{HengEt2016ApJL}
{Heng}, K. 2016, \apjl, 826, L16

\bibitem[{{Henry} {et~al.}(2006){Henry}, {Jao}, {Subasavage}, {Beaulieu},
  {Ianna}, {Costa}, \& {M{\'e}ndez}}]{HenryEt2006AJ}
{Henry}, T.~J., {Jao}, W.-C., {Subasavage}, J.~P., {et~al.} 2006, \aj, 132,
  2360

\bibitem[{{Holton} {et~al.}(1995){Holton}, {Haynes}, {McIntyre}, {Douglass},
  {Rood}, \& {Pfister}}]{HoltonEt1995}
{Holton}, J.~R., {Haynes}, P.~H., {McIntyre}, M.~E., {et~al.} 1995, Reviews of
  Geophysics, 33, 403

\bibitem[{Horton {et~al.}(2012)Horton, Poulsen, Monta{\~n}ez, \&
  DiMichele}]{HortonEt2012PPP}
Horton, D.~E., Poulsen, C.~J., Monta{\~n}ez, I.~P., \& DiMichele, W.~A. 2012,
  Palaeogeography, Palaeoclimatology, Palaeoecology, 331, 150

\bibitem[{{Hu} {et~al.}(2012){Hu}, {Seager}, \& {Bains}}]{HuEt2012ApJ}
{Hu}, R., {Seager}, S., \& {Bains}, W. 2012, \apj, 761, 166

\bibitem[{{Hunten}(1973)}]{Hunten1973JAS}
{Hunten}, D.~M. 1973, Journal of Atmospheric Sciences, 30, 1481

\bibitem[{{Husser} {et~al.}(2013){Husser}, {Wende-von Berg}, {Dreizler},
  {Homeier}, {Reiners}, {Barman}, \& {Hauschildt}}]{HusserEt2013A&A}
{Husser}, T.-O., {Wende-von Berg}, S., {Dreizler}, S., {et~al.} 2013, Astronomy
  \& Astrophysics, 553, A6

\bibitem[{{Joshi} {et~al.}(1997){Joshi}, {Haberle}, \&
  {Reynolds}}]{JoshiEt1997Icarus}
{Joshi}, M.~M., {Haberle}, R.~M., \& {Reynolds}, R.~T. 1997, Icarus, 129, 450

\bibitem[{{Kaltenegger} \& {Sasselov}(2010)}]{Kaltenegger+2010ApJ}
{Kaltenegger}, L., \& {Sasselov}, D. 2010, \apj, 708, 1162

\bibitem[{{Kasting}(1988)}]{Kasting1988Icarus}
{Kasting}, J.~F. 1988, \icarus, 74, 472

\bibitem[{{Kasting} {et~al.}(2015){Kasting}, {Chen}, \&
  {Kopparapu}}]{KastingEt2015ApJL}
{Kasting}, J.~F., {Chen}, H., \& {Kopparapu}, R.~K. 2015, \apjl, 813, L3

\bibitem[{{Kasting} {et~al.}(1984){Kasting}, {Pollack}, \&
  {Ackerman}}]{KastingEt1984Icarus}
{Kasting}, J.~F., {Pollack}, J.~B., \& {Ackerman}, T.~P. 1984, Icarus, 57, 335

\bibitem[{{Kasting} {et~al.}(1993){Kasting}, {Whitmire}, \&
  {Reynolds}}]{KastingEt1993Icarus}
{Kasting}, J.~F., {Whitmire}, D.~P., \& {Reynolds}, R.~T. 1993, Icarus, 101,
  108

\bibitem[{{Kawashima} \& {Ikoma}(2018)}]{Kawashima+Ikoma2018ApJ}
{Kawashima}, Y., \& {Ikoma}, M. 2018, \apj, 853, 7

\bibitem[{{Kempton} {et~al.}(2017){Kempton}, {Bean}, \&
  {Parmentier}}]{KemptonEt2017ApJL}
{Kempton}, E.~M.-R., {Bean}, J.~L., \& {Parmentier}, V. 2017, \apjl, 845, L20

\bibitem[{Kiehl \& Ramanathan(1983)}]{kiehl1983co2}
Kiehl, J., \& Ramanathan, V. 1983, Journal of Geophysical Research: Oceans, 88,
  5191

\bibitem[{{Kilic} {et~al.}(2017){Kilic}, {Raible}, \&
  {Stocker}}]{KilicEt2017ApJ}
{Kilic}, C., {Raible}, C.~C., \& {Stocker}, T.~F. 2017, \apj, 844, 147

\bibitem[{{Kinnison} {et~al.}(2007){Kinnison}, {Brasseur}, {Walters}, {Garcia},
  {Marsh}, {Sassi}, {Harvey}, {Randall}, {Emmons}, {Lamarque}, {Hess},
  {Orlando}, {Tie}, {Randel}, {Pan}, {Gettelman}, {Granier}, {Diehl},
  {Niemeier}, \& {Simmons}}]{KinnisonEt2007JGR}
{Kinnison}, D.~E., {Brasseur}, G.~P., {Walters}, S., {et~al.} 2007, Journal of
  Geophysical Research (Atmospheres), 112, D20302

\bibitem[{{Kodama} {et~al.}(2018){Kodama}, {Nitta}, {Genda}, {Takao}, {O'ishi},
  {Abe-Ouchi}, \& {Abe}}]{KodamaEt2018JGR}
{Kodama}, T., {Nitta}, A., {Genda}, H., {et~al.} 2018, Journal of Geophysical
  Research (Planets), 123, 559

\bibitem[{{Koll} {et~al.}(2019){Koll}, {Malik}, {Mansfield}, {Kempton}, {Kite},
  {Abbot}, \& {Bean}}]{KollEt2019arXiv}
{Koll}, D.~D.~B., {Malik}, M., {Mansfield}, M., {et~al.} 2019, arXiv e-prints,
  arXiv:1907.13138

\bibitem[{{Komacek} \& {Abbot}(2016)}]{Komacek+Abbot2016ApJ}
{Komacek}, T.~D., \& {Abbot}, D.~S. 2016, \apj, 832, 54

\bibitem[{{Komacek} \& {Abbot}(2019)}]{Komacek+Abbot2019ApJ}
---. 2019, \apj, 871, 245

\bibitem[{{Kopparapu} {et~al.}(2017){Kopparapu}, {Wolf}, {Arney}, {Batalha},
  {Haqq-Misra}, {Grimm}, \& {Heng}}]{KopparapuEt2017ApJ}
{Kopparapu}, R.~k., {Wolf}, E.~T., {Arney}, G., {et~al.} 2017, The
  Astrophysical Journal, 845, 5

\bibitem[{{Kopparapu} {et~al.}(2016){Kopparapu}, {Wolf}, {Haqq-Misra}, {Yang},
  {Kasting}, {Meadows}, {Terrien}, \& {Mahadevan}}]{KopparapuEt2016ApJ}
{Kopparapu}, R.~k., {Wolf}, E.~T., {Haqq-Misra}, J., {et~al.} 2016, The
  Astrophysical Journal, 819, 84

\bibitem[{{Kopparapu} {et~al.}(2013){Kopparapu}, {Ramirez}, {Kasting}, {Eymet},
  {Robinson}, {Mahadevan}, {Terrien}, {Domagal-Goldman}, {Meadows}, \&
  {Deshpande}}]{KopparapuEt2013ApJ}
{Kopparapu}, R.~K., {Ramirez}, R., {Kasting}, J.~F., {et~al.} 2013, The
  Astrophysical Journal, 765, 131

\bibitem[{{Kozakis} {et~al.}(2018){Kozakis}, {Kaltenegger}, \&
  {Hoard}}]{KozakisEt2018ApJ}
{Kozakis}, T., {Kaltenegger}, L., \& {Hoard}, D.~W. 2018, \apj, 862, 69

\bibitem[{{Lean}(2000)}]{Lean2000GRL}
{Lean}, J. 2000, \grl, 27, 2425

\bibitem[{Lean {et~al.}(1995)Lean, Beer, \& Bradley}]{lean1995reconstruction}
Lean, J., Beer, J., \& Bradley, R. 1995, Geophysical Research Letters, 22, 3195

\bibitem[{{Leconte} {et~al.}(2013{\natexlab{a}}){Leconte}, {Forget}, {Charnay},
  {Wordsworth}, \& {Pottier}}]{LeconteEt2013NATURE}
{Leconte}, J., {Forget}, F., {Charnay}, B., {Wordsworth}, R., \& {Pottier}, A.
  2013{\natexlab{a}}, \nat, 504, 268

\bibitem[{{Leconte} {et~al.}(2013{\natexlab{b}}){Leconte}, {Forget}, {Charnay},
  {Wordsworth}, {Selsis}, {Millour}, \& {Spiga}}]{LeconteEt2013A&A}
{Leconte}, J., {Forget}, F., {Charnay}, B., {et~al.} 2013{\natexlab{b}}, \aap,
  554, A69

\bibitem[{{Leconte} {et~al.}(2015){Leconte}, {Wu}, {Menou}, \&
  {Murray}}]{LeconteEt2015SCI}
{Leconte}, J., {Wu}, H., {Menou}, K., \& {Murray}, N. 2015, Science, 347, 632

\bibitem[{{Lehmer} {et~al.}(2018){Lehmer}, {Catling}, {Parenteau}, \&
  {Hoehler}}]{LehmerEt2018ApJ}
{Lehmer}, O.~R., {Catling}, D.~C., {Parenteau}, M.~N., \& {Hoehler}, T.~M.
  2018, \apj, 859, 171

\bibitem[{{Lewis} {et~al.}(2018){Lewis}, {Lambert}, {Boutle}, {Mayne},
  {Manners}, \& {Acreman}}]{LewisEt2018ApJ}
{Lewis}, N.~T., {Lambert}, F.~H., {Boutle}, I.~A., {et~al.} 2018, \apj, 854,
  171

\bibitem[{{Lincowski} {et~al.}(2018){Lincowski}, {Meadows}, {Crisp},
  {Robinson}, {Luger}, {Lustig-Yaeger}, \& {Arney}}]{LincowskiEt2018ApJ}
{Lincowski}, A.~P., {Meadows}, V.~S., {Crisp}, D., {et~al.} 2018, \apj, 867, 76

\bibitem[{{Lingam} \& {Loeb}(2017)}]{Lingam+Loeb2017ApJL}
{Lingam}, M., \& {Loeb}, A. 2017, \apjl, 846, L21

\bibitem[{{Lingam} \& {Loeb}(2018)}]{Lingam+Loeb2018JCAP}
---. 2018, \jcap, 5, 020

\bibitem[{{Lingam} \& {Loeb}(2019{\natexlab{a}})}]{Lingam+Loeb2019AJ}
---. 2019{\natexlab{a}}, \aj, 157, 25

\bibitem[{{Lingam} \& {Loeb}(2019{\natexlab{b}})}]{LingamEt2019MNRAS}
---. 2019{\natexlab{b}}, \mnras, 485, 5924

\bibitem[{{Loyd} {et~al.}(2016){Loyd}, {France}, {Youngblood}, {Schneider},
  {Brown}, {Hu}, {Linsky}, {Froning}, {Redfield}, {Rugheimer}, \&
  {Tian}}]{LoydEt2016ApJ}
{Loyd}, R.~O.~P., {France}, K., {Youngblood}, A., {et~al.} 2016, \apj, 824, 102

\bibitem[{{Luger} \& {Barnes}(2015)}]{Luger+Barnes2015AsBio}
{Luger}, R., \& {Barnes}, R. 2015, Astrobiology, 15, 119

\bibitem[{{Malik} {et~al.}(2019){Malik}, {Kempton}, {Koll}, {Mansfield},
  {Bean}, \& {Kite}}]{MalikEt2019arXiv}
{Malik}, M., {Kempton}, E.~M.-R., {Koll}, D.~D.~B., {et~al.} 2019, arXiv
  e-prints, arXiv:1907.13135

\bibitem[{{Marsh} {et~al.}(2013){Marsh}, {Mills}, {Kinnison}, {Lamarque},
  {Calvo}, \& {Polvani}}]{MarshEt2013JGR}
{Marsh}, D.~R., {Mills}, M.~J., {Kinnison}, D.~E., {et~al.} 2013, Journal of
  Climate, 26, 7372

\bibitem[{Matsuno(1966)}]{Matsuno1966}
Matsuno, T. 1966, Journal of the Meteorological Society of Japan. Ser. II, 44,
  25

\bibitem[{{Merlis} \& {Schneider}(2010)}]{Merlis+Schneider2010}
{Merlis}, T.~M., \& {Schneider}, T. 2010, Journal of Advances in Modeling Earth
  Systems, 2, 13

\bibitem[{{Miller-Ricci} {et~al.}(2009){Miller-Ricci}, {Meyer}, {Seager}, \&
  {Elkins-Tanton}}]{KemptonEt2009ApJ}
{Miller-Ricci}, E., {Meyer}, M.~R., {Seager}, S., \& {Elkins-Tanton}, L. 2009,
  \apj, 704, 770

\bibitem[{{Morley} {et~al.}(2013){Morley}, {Fortney}, {Kempton}, {Marley},
  {Visscher}, \& {Zahnle}}]{MorleyEt2013ApJ}
{Morley}, C.~V., {Fortney}, J.~J., {Kempton}, E.~M.-R., {et~al.} 2013, \apj,
  775, 33

\bibitem[{Nakajima {et~al.}(1992)Nakajima, Hayashi, \& Abe}]{NakajimaEt1992JRG}
Nakajima, S., Hayashi, Y.-Y., \& Abe, Y. 1992, Journal of the Atmospheric
  Sciences, 49, 2256

\bibitem[{Neale {et~al.}(2010)Neale, Chen, Gettelman, Lauritzen, Park,
  Williamson, Conley, Garcia, Kinnison, Lamarque,
  {et~al.}}]{neale2010description}
Neale, R.~B., Chen, C.-C., Gettelman, A., {et~al.} 2010, NCAR Tech. Note
  NCAR/TN-486+ STR, 1, 1

\bibitem[{{Olson} {et~al.}(2018){Olson}, {Schwieterman}, {Reinhard},
  {Ridgwell}, {Kane}, {Meadows}, \& {Lyons}}]{OlsonEt2018ApJL}
{Olson}, S.~L., {Schwieterman}, E.~W., {Reinhard}, C.~T., {et~al.} 2018, \apjl,
  858, L14

\bibitem[{{O'Malley-James} \&
  {Kaltenegger}(2017)}]{OMalley+Kaltenegger2017MNRAS}
{O'Malley-James}, J.~T., \& {Kaltenegger}, L. 2017, \mnras, 469, L26

\bibitem[{{Paradise} \& {Menou}(2017)}]{Paradise+Menou2017ApJ}
{Paradise}, A., \& {Menou}, K. 2017, \apj, 848, 33

\bibitem[{{Ramirez} \& {Kaltenegger}(2017)}]{Ramirez+Kaltenegger2017ApJL}
{Ramirez}, R.~M., \& {Kaltenegger}, L. 2017, \apjl, 837, L4

\bibitem[{{Ranjan} {et~al.}(2017){Ranjan}, {Wordsworth}, \&
  {Sasselov}}]{RanjanEt2017ApJ}
{Ranjan}, S., {Wordsworth}, R., \& {Sasselov}, D.~D. 2017, \apj, 843, 110

\bibitem[{{Raymond} {et~al.}(2006){Raymond}, {Quinn}, \&
  {Lunine}}]{RaymondEt2006Icarus}
{Raymond}, S.~N., {Quinn}, T., \& {Lunine}, J.~I. 2006, \icarus, 183, 265

\bibitem[{{Roble} \& {Ridley}(1994)}]{RobleEt1994GRL}
{Roble}, R.~G., \& {Ridley}, E.~C. 1994, \grl, 21, 417

\bibitem[{{Romps} \& {Kuang}(2009)}]{Romps+Kuang2009GRL}
{Romps}, D.~M., \& {Kuang}, Z. 2009, \grl, 36, L09804

\bibitem[{{Rugheimer} {et~al.}(2015){Rugheimer}, {Kaltenegger}, {Segura},
  {Linsky}, \& {Mohanty}}]{RugheimerEt2015ApJ}
{Rugheimer}, S., {Kaltenegger}, L., {Segura}, A., {Linsky}, J., \& {Mohanty},
  S. 2015, \apj, 809, 57

\bibitem[{{Sagan} {et~al.}(1993){Sagan}, {Thompson}, {Carlson}, {Gurnett}, \&
  {Hord}}]{SaganEt1993NATURE}
{Sagan}, C., {Thompson}, W.~R., {Carlson}, R., {Gurnett}, D., \& {Hord}, C.
  1993, Nature, 365, 715

\bibitem[{{Schwieterman} {et~al.}(2019){Schwieterman}, {Reinhard}, {Olson},
  {Harman}, \& {Lyons}}]{SchwietermanEt2019ApJ}
{Schwieterman}, E.~W., {Reinhard}, C.~T., {Olson}, S.~L., {Harman}, C.~E., \&
  {Lyons}, T.~W. 2019, \apj, 878, 19

\bibitem[{{Segura} {et~al.}(2005){Segura}, {Kasting}, {Meadows}, {Cohen},
  {Scalo}, {Crisp}, {Butler}, \& {Tinetti}}]{SeguraEt2005AsBio}
{Segura}, A., {Kasting}, J.~F., {Meadows}, V., {et~al.} 2005, Astrobiology, 5,
  706

\bibitem[{{Shen} \& {Turner}(2008)}]{ShenEt2008ApJ}
{Shen}, Y., \& {Turner}, E.~L. 2008, \apj, 685, 553

\bibitem[{{Showman} \& {Polvani}(2011)}]{Showman+Polvani2011ApJ}
{Showman}, A.~P., \& {Polvani}, L.~M. 2011, \apj, 738, 71

\bibitem[{{Solomon} \& {Qian}(2005)}]{Solomon+Qian2005JGR}
{Solomon}, S.~C., \& {Qian}, L. 2005, Journal of Geophysical Research (Space
  Physics), 110, A10306

\bibitem[{{Spinelli} {et~al.}(2019){Spinelli}, {Borsa}, {Ghirlanda},
  {Ghisellini}, {Campana}, {Haardt}, \& {Poretti}}]{SpenelliEt2019arXiv}
{Spinelli}, R., {Borsa}, F., {Ghirlanda}, G., {et~al.} 2019, arXiv e-prints,
  arXiv:1906.08783

\bibitem[{{Stevenson}(2016)}]{StevensonEt2016ApJL}
{Stevenson}, K.~B. 2016, \apjl, 817, L16

\bibitem[{{Tarter} {et~al.}(2007){Tarter}, {Backus}, {Mancinelli}, {Aurnou},
  {Backman}, {Basri}, {Boss}, {Clarke}, {Deming}, {Doyle}, {Feigelson},
  {Freund}, {Grinspoon}, {Haberle}, {Hauck}, {Heath}, {Henry}, {Hollingsworth},
  {Joshi}, {Kilston}, {Liu}, {Meikle}, {Reid}, {Rothschild}, {Scalo}, {Segura},
  {Tang}, {Tiedje}, {Turnbull}, {Walkowicz}, {Weber}, \&
  {Young}}]{TarterEt2007}
{Tarter}, J.~C., {Backus}, P.~R., {Mancinelli}, R.~L., {et~al.} 2007,
  Astrobiology, 7, 30

\bibitem[{{Tian} \& {Ida}(2015)}]{Tian+Ida2015NatGeo}
{Tian}, F., \& {Ida}, S. 2015, Nature Geoscience, 8, 177

\bibitem[{{Tsiganis} {et~al.}(2005){Tsiganis}, {Gomes}, {Morbidelli}, \&
  {Levison}}]{TsiganisEt2005NAT}
{Tsiganis}, K., {Gomes}, R., {Morbidelli}, A., \& {Levison}, H.~F. 2005, \nat,
  435, 459

\bibitem[{{Vanderspek} {et~al.}(2019){Vanderspek}, {Huang}, {Vanderburg},
  {Ricker}, {Latham}, {Seager}, {Winn}, {Jenkins}, {Burt}, {Dittmann},
  {Newton}, {Quinn}, {Shporer}, {Charbonneau}, {Irwin}, {Ment}, {Winters},
  {Collins}, {Evans}, {Gan}, {Hart}, {Jensen}, {Kielkopf}, {Mao}, {Waalkes},
  {Bouchy}, {Marmier}, {Nielsen}, {Ottoni}, {Pepe}, {S{\'e}gransan}, {Udry},
  {Henry}, {Paredes}, {James}, {Hinojosa}, {Silverstein}, {Palle},
  {Berta-Thompson}, {Crossfield}, {Davies}, {Dragomir}, {Fausnaugh}, {Glidden},
  {Pepper}, {Morgan}, {Rose}, {Twicken}, {Villase{\~n}or}, {Yu}, {Bakos},
  {Bean}, {Buchhave}, {Christensen-Dalsgaard}, {Christiansen}, {Ciardi},
  {Clampin}, {De Lee}, {Deming}, {Doty}, {Jernigan}, {Kaltenegger}, {Lissauer},
  {McCullough}, {Narita}, {Paegert}, {Pal}, {Rinehart}, {Sasselov}, {Sato},
  {Sozzetti}, {Stassun}, \& {Torres}}]{VanderspekEt2019ApJL}
{Vanderspek}, R., {Huang}, C.~X., {Vanderburg}, A., {et~al.} 2019, \apjl, 871,
  L24

\bibitem[{{Vladilo} {et~al.}(2013){Vladilo}, {Murante}, {Silva}, {Provenzale},
  {Ferri}, \& {Ragazzini}}]{VladiloEt2013ApJ}
{Vladilo}, G., {Murante}, G., {Silva}, L., {et~al.} 2013, \apj, 767, 65

\bibitem[{Wakeford {et~al.}(2019)Wakeford, Wilson, Stevenson, \&
  Lewis}]{WakefordEt2019arXiv}
Wakeford, H., Wilson, T., Stevenson, K., \& Lewis, N. 2019, arXiv preprint
  arXiv:1908.10669

\bibitem[{{Wakeford} {et~al.}(2017){Wakeford}, {Sing}, {Kataria}, {Deming},
  {Nikolov}, {Lopez}, {Tremblin}, {Amundsen}, {Lewis}, {Mandell}, {Fortney},
  {Knutson}, {Benneke}, \& {Evans}}]{WakefordEt2017SCI}
{Wakeford}, H.~R., {Sing}, D.~K., {Kataria}, T., {et~al.} 2017, Science, 356,
  628

\bibitem[{{Way} {et~al.}(2018){Way}, {Del Genio}, {Aleinov}, {Clune}, {Kelley},
  \& {Kiang}}]{WayEt2018ApJS}
{Way}, M.~J., {Del Genio}, A.~D., {Aleinov}, I., {et~al.} 2018, \apjs, 239, 24

\bibitem[{{Way} {et~al.}(2016){Way}, {Del Genio}, {Kiang}, {Sohl}, {Grinspoon},
  {Aleinov}, {Kelley}, \& {Clune}}]{WayEt2016GRL}
{Way}, M.~J., {Del Genio}, A.~D., {Kiang}, N.~Y., {et~al.} 2016, \grl, 43, 8376

\bibitem[{{Wolf} {et~al.}(2019){Wolf}, {Kopparapu}, \&
  {Haqq-Misra}}]{WolfEt2019ApJ}
{Wolf}, E.~T., {Kopparapu}, R.~K., \& {Haqq-Misra}, J. 2019, \apj, 877, 35

\bibitem[{{Wolf} \& {Toon}(2013)}]{Wolf+Toon2013Asbio}
{Wolf}, E.~T., \& {Toon}, O.~B. 2013, Astrobiology, 13, 656

\bibitem[{{Wolf} \& {Toon}(2015)}]{Wolf+Toon2015JGR}
---. 2015, Journal of Geophysical Research (Atmospheres), 120, 5775

\bibitem[{{Yang} {et~al.}(2019{\natexlab{a}}){Yang}, {Abbot}, {Koll}, {Hu}, \&
  {Showman}}]{YangEt2019ApJa}
{Yang}, J., {Abbot}, D.~S., {Koll}, D.~D.~B., {Hu}, Y., \& {Showman}, A.~P.
  2019{\natexlab{a}}, \apj, 871, 29

\bibitem[{{Yang} {et~al.}(2014){Yang}, {Bou{\'e}}, {Fabrycky}, \&
  {Abbot}}]{YangEt2014ApJL}
{Yang}, J., {Bou{\'e}}, G., {Fabrycky}, D.~C., \& {Abbot}, D.~S. 2014, \apjl,
  787, L2

\bibitem[{{Yang} {et~al.}(2013){Yang}, {Cowan}, \& {Abbot}}]{YangEt2013ApJL}
{Yang}, J., {Cowan}, N.~B., \& {Abbot}, D.~S. 2013, \apjl, 771, L45

\bibitem[{{Yang} {et~al.}(2019{\natexlab{b}}){Yang}, {Leconte}, {Wolf},
  {Merlis}, {Koll}, {Forget}, \& {Abbot}}]{YangEt2019ApJb}
{Yang}, J., {Leconte}, J., {Wolf}, E.~T., {et~al.} 2019{\natexlab{b}}, \apj,
  875, 46

\bibitem[{{Yang} {et~al.}(2016){Yang}, {Leconte}, {Wolf}, {Goldblatt}, {Feldl},
  {Merlis}, {Wang}, {Koll}, {Ding}, {Forget}, \& {Abbot}}]{YangEt2016ApJ}
---. 2016, \apj, 826, 222

\bibitem[{{Youngblood} {et~al.}(2016){Youngblood}, {France}, {Loyd}, {Linsky},
  {Redfield}, {Schneider}, {Wood}, {Brown}, {Froning}, {Miguel}, {Rugheimer},
  \& {Walkowicz}}]{YoungbloodEt2016ApJ}
{Youngblood}, A., {France}, K., {Loyd}, R.~O.~P., {et~al.} 2016, \apj, 824, 101

\bibitem[{Zhang \& McFarlane(1995)}]{zhang1995sensitivity}
Zhang, G.~J., \& McFarlane, N.~A. 1995, Atmosphere-ocean, 33, 407

\bibitem[{Zhang {et~al.}(2003)Zhang, Lin, Bretherton, Hack, \&
  Rasch}]{zhang2003modified}
Zhang, M., Lin, W., Bretherton, C.~S., Hack, J.~J., \& Rasch, P.~J. 2003,
  Journal of Geophysical Research: Atmospheres, 108, ACL

\bibitem[{{Zhang} \& {Showman}(2018)}]{Zhang&Showman2018ApJ}
{Zhang}, X., \& {Showman}, A.~P. 2018, \apj, 866, 1

\bibitem[{{Zsom} {et~al.}(2013){Zsom}, {Seager}, {de Wit}, \&
  {Stamenkovi{\'c}}}]{ZsomEt2013ApJ}
{Zsom}, A., {Seager}, S., {de Wit}, J., \& {Stamenkovi{\'c}}, V. 2013, \apj,
  778, 109

\end{thebibliography}

\end{document}